\documentclass[twocolumn]{aastex631}

\usepackage{csquotes} 
\usepackage{xspace}
\usepackage{xcolor}
\usepackage{amsmath} 

\newcommand{\AL}{$A_L$\xspace}
\newcommand{\ALw}{$A_{L, w}$\xspace}
\newcommand{\ALuw}{$A_{L, uw}$\xspace}
\newcommand{\DAL}{$\Delta A_L$\xspace}
\newcommand{\DALAL}{$\Delta A_L/A_{L, w}$\xspace}

\graphicspath{{./}}

\shorttitle{Alignment Parameters}
\shortauthors{Chen et al.}

\begin{document}
	
\title{Alignment Parameters: Quantifying Dense Core Alignment in Star-forming Regions}

\author[0000-0002-8336-2837]{Wei-An Chen}
\affiliation{Graduate Institute of Astrophysics, National Taiwan University, No.\ 1, Sec.\ 4, Roosevelt Rd., Taipei 10617, Taiwan, R.O.C}
\affiliation{Institute of Astronomy and Astrophysics, Academia Sinica, 11F of Astronomy-Mathematics Building, No.\ 1, Sec.\ 4, Roosevelt Rd, Taipei 106216, Taiwan, R.O.C.}
\email{wachen@asiaa.sinica.edu.tw}

\author[0000-0002-0675-276X]{Ya-Wen Tang}
\affiliation{Institute of Astronomy and Astrophysics, Academia Sinica, 11F of Astronomy-Mathematics Building, No.\ 1, Sec.\ 4, Roosevelt Rd, Taipei 106216, Taiwan, R.O.C.}

\author[0000-0001-9751-4603]{S. D. Clarke}
\affiliation{Institute of Astronomy and Astrophysics, Academia Sinica, 11F of Astronomy-Mathematics Building, No.\ 1, Sec.\ 4, Roosevelt Rd, Taipei 106216, Taiwan, R.O.C.}

\author[0000-0002-7125-7685]{Patricio Sanhueza}
\affiliation{Department of Earth and Planetary Sciences, Tokyo Institute of Technology, Meguro, Tokyo, 152-8551, Japan}
\affiliation{National Astronomical Observatory of Japan, National Institutes of Natural Sciences, 2-21-1 Osawa, Mitaka, Tokyo 181-8588, Japan}

\begin{abstract}
	Recent high-resolution observations at millimeter (mm) and sub-mm reveal a diverse spatial distribution for sub-pc scale dense cores within star-forming regions, ranging from clustered to aligned arrangements. 
	To address the increasing volume of observational and simulation data, we introduce \textquote{alignment parameters} as a quantitative and reproducible method to automatically assess core alignment. 
	We first demonstrate the effectiveness of these parameters by applying them to artificial test clumps and comparing the results with labels from visual inspection. 
	A threshold value is then proposed to differentiate between \textquote{clustered} and \textquote{aligned} categories. 
	Subsequently, we apply these parameters to dense cores identified from a sample of ALMA 1.3 mm dust continuum images in high-mass star-forming regions. 
	Analysis exploring correlations between alignment parameters and clump properties rules out the presence of moderate or strong correlation, indicating that clump properties do not appear to strongly influence the outcome of fragmentation. One possible explanation for this is that the fragmentation process is chaotic, meaning that small variations in initial conditions can lead to significant differences in fragmentation outcomes, thus obscuring any direct link between clump properties and core alignment/distribution.
\end{abstract}


\section{Introduction} \label{sec:intro}
High-mass stars, typically exceeding 8 $M_{\odot}$, form within clusters and originate from dense, self-gravitating cores at sub-pc scales \citep{2003ARA&A..41...57L}. 
These cores are hierarchically embedded within larger structures: pc-scale dense clumps and even more extensive molecular clouds spanning several parsecs, all of which reside within even larger warm interstellar medium (ISM) (see review in \citealp{2023ASPC..534..233P}). 
The process by which molecular clouds gravitationally collapse and fragment into substructures, such as clumps and cores, is known as fragmentation. 
A complex interplay of gravity, magnetic fields, and turbulence plays a significant role in the fragmentation and formation of dense cores. 
Additional factors, such as large-scale compression, nearby stellar feedback, and the morphology of the parent structures, can further shape this process. 
This complex interplay likely contributes to the observed diversity in the distribution and properties of dense cores at sub-pc scales. 
Given such potential connection between fragmentation and the interplay of physical mechanisms, distributions and properties of dense cores can serve as a diagnostic tool to probe the dynamics of their host clumps (e.g., \citealp{2015MNRAS.453.3785P, 2018A&A...617A.100B, 2019A&A...632A..83S, 2019ApJ...878...10T, 2019ApJ...886..102S, 2020ApJ...895..142L, 2021ApJ...912..159P, 2021A&A...649A.113B, 2021ApJ...912L..27E, 2022AJ....164..175C, 2023MNRAS.526.2278A, 2023ApJ...951...68C, 2023ApJ...950..148M, 2024ApJS..270....9X, 2024ApJ...963..126G, 2024ApJ...966..171M, 2024ApJ...974...95I}).

Building upon the established link between core distributions and clump properties, several observational studies have been conducted. 
\cite{2019ApJ...878...10T} investigated the G34.43+00.24 region, visually identifying three distinct core alignment patterns across its three clumps (MM1, MM2, and MM3): \textit{no fragmentation}, \textit{aligned fragmentation}, and \textit{clustered fragmentation}. 
Their analysis suggested that the relative energetic importance of gravity, turbulence, and magnetic fields might underlie these distinct fragmentation patterns. 
Aligned fragmentation is characterized by cores placed along a predominant direction under magnetic field dominance, while clustered fragmentation exhibits cores distributed randomly without a clear directional preference when turbulence in the regions cannot be ignored. 
No fragmentation is observed when a single dominant core exists without companions, which occurs when factors other than gravity are negligible. 
Subsequently, \cite{2022AJ....164..175C, 2023ApJ...951...68C} adopted these three fragmentation categories, reporting a consistent trend linking core alignment to the dominance of different energy components.

However, the question of whether fragmentation consistently yields aligned or clustered structures solely due to the dominance of specific energy components remains unresolved. 
Recent work by Lee et al. (in prep.) reported a lack of fragmentation in regions with strong magnetic fields and weak turbulence, consistent with the known suppression of fragmentation by magnetic fields \citep{2011ApJ...742L...9C, 2013ApJ...779...96T, 2017ApJ...848....2H, 2018A&A...614A..64B}. 
This raises the question of how strong magnetic fields must be to sustain aligned fragmentation.
Additionally, anisotropic collapse of a structure can result in filamentary geometry, where dense cores may form due to varying collapse rates along different axes, potentially leading to highly aligned configurations \citep{2016MNRAS.458..319C, 2017MNRAS.468.2489C, 2020MNRAS.497.4390C, 2016MNRAS.463.4301H, 2018MNRAS.481L...1H, 2015MNRAS.452.2410S, 2017ApJ...848....2H, 2019ApJ...881...97H}.
Furthermore, turbulence and substructure within the larger structure can influence the core formation process, resulting in cores being associated with the presence of fibers and sub-filaments \citep{2015A&A...574A.104T, 2017A&A...606A.123H, 2018A&A...610A..77H, 2017MNRAS.468.2489C, 2018MNRAS.479.1722C, 2020MNRAS.497.4390C}. 
Several studies also indicate that core separation decreases and core distributions become more compact in more evolved clumps \citep{2018A&A...617A.100B, 2021A&A...649A.113B, 2023MNRAS.520.2306T, 2024ApJS..270....9X, 2024ApJ...974...95I}. 
Collectively, these findings suggest a dynamic and complex fragmentation process within star-forming clumps.

The increasing volume of data from both observational surveys and state-of-the-art simulations is crucial for unraveling these questions and advancing our understanding of high-mass star formation, particularly the processes by which dense cores form under diverse environmental conditions. 
To effectively study the relation between core distributions and clump properties, a robust metric is needed to automatically quantify core alignment and systematically investigate its relationship with larger-scale structures. 
This paper introduces \textquote{alignment parameters} as a method to quantify core alignment. 
We demonstrate the effectiveness of these parameters by applying them to 1.3 mm dust continuum data from the ALMA Survey of 70 $\mu$m Dark High-mass Clumps in Early Stages (ASHES; \citealp{2019ApJ...886..102S, 2023ApJ...950..148M}) to quantify core alignment within these clumps. 
Subsequently, we explore potential correlations between core alignment and clump-scale properties derived from the ASHES survey.

This paper is structured as follows. 
Section~\ref{sec:DevelopeAL} outlines the development of the alignment parameters and their subsequent validation. 
Section~\ref{sec:CorrObs} applies these parameters to the dust continuum images from the ASHES survey and discusses the resulting correlations. 
Finally, Section~\ref{sec:Conclusion} summarizes the key findings and results.

\section{The Alignment Parameter, \AL} \label{sec:DevelopeAL}
\begin{figure}[htb!]
	\epsscale{1.1}
	\plotone{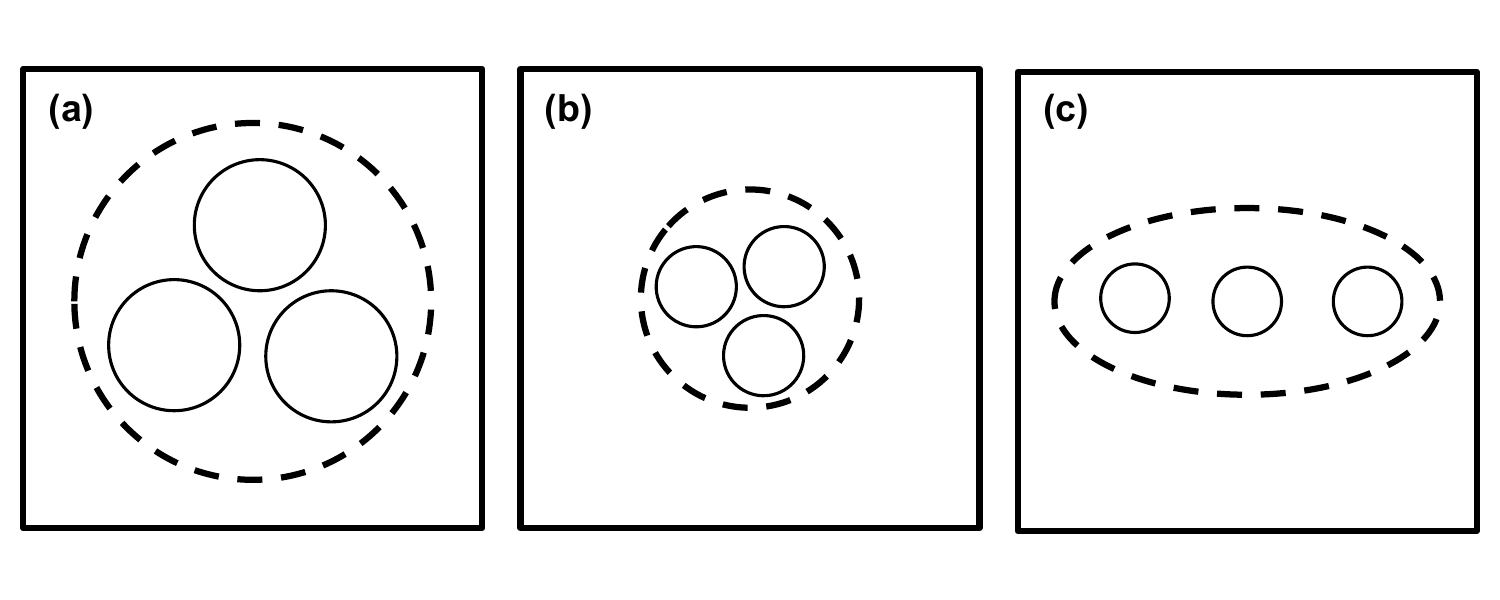}
	\caption{Examples for quantifying core alignment (introduced in Section~\ref{sec:DevelopeAL}). Panels (a) and (b) depict clumps with similar core configurations (solid and dashed lines represent clump and core boundaries, respectively) but different sizes. Panel (c) illustrates a well-aligned core configuration.}
	\label{fig:ALEx}
\end{figure}

Core positions are the main parameter for quantifying core alignment. 
These positions can be identified using various structure-finding algorithms, such as dendrograms \citep{Rosolowsky2008}. 
However, a challenge arises when using absolute core positions and separations to compare clumps of different sizes but exhibiting similar core arrangements.
Figure~\ref{fig:ALEx} illustrates this issue. 
Panels (a) and (b) depict clumps with similar core configurations (solid and dashed lines represent clump and core boundaries, respectively) but different sizes. 
To address this point, we introduce a normalization step for the core separations ($S_{ij}$) between each core pair. 
This normalization is achieved by 
\begin{equation} \label{equ:Sij}
	S'_{ij} = \frac{S_{ij}}{\sigma_m},
\end{equation}
where $\sigma_m$ is the maximum value between the size of the beam's minor axis and the characteristic minor axis length obtained through (weighted) principal component analysis (PCA). 
The use of the size of the beam's minor axis accounts for the inherent ambiguity in separation measurements when the minor axis length is small.
To ensure finite values for the relative core separations ($S'_{ij}$), alignment parameters are calculated only for clumps with more than two cores.

This normalization step allows for a direct comparison of $S'_{ij}$, highlighting the varying degrees of core alignment. 
In a well-aligned core configuration (Figure~\ref{fig:ALEx} (c)), we expect a higher number of core pairs with $S'_{ij} > 1$ compared to a more clustered scenario (Figure~\ref{fig:ALEx} (a) and (b)), where $S'_{ij}$ values will be closer to or less than 1. 
Notably, since the core alignments in Figure~\ref{fig:ALEx} (a) and (b) are similar, their corresponding $S'_{ij}$ distributions are expected to be comparable.

To represent the overall core alignment, we calculate the mean value of $S'_{ij}$. 
For investigations focused on the overall core alignment without considering specific core properties, the arithmetic mean provides a suitable measure. 
This \textit{unweighted alignment parameter} (\ALuw) is then defined as:
\begin{equation} \label{eqn:ALuw}
	A_{L, uw} = \text{mean}(S'_{ij}).
\end{equation}
Alternatively, the mean value can be weighted to emphasize the alignment of specific core properties relevant to the scientific question. 
For example, weights could be assigned based on core flux or mass. 
The resulting \textit{weighted alignment parameter} (\ALw) is defined as:
\begin{equation} \label{eqn:ALw}
	A_{L, w} = \frac{\sum_{i \neq j}^N w_i w_j S'_{ij}}{{\sum_{i \neq j}^N w_i w_j} },
\end{equation}
where $w_i$ represents the weight assigned to the $i^{th}$ core and $N$ is the total number of cores.
In this case, the same weights are used in weighted PCA to obtain $\sigma_m$.
The uncertainty associated with both \ALuw and \ALw can be estimated from the error of the (weighted) mean.

Once we incorporate the core mass or flux as weights in the alignment parameter calculation, it  allows us to investigate differences in the alignment between massive (brighter) and less massive (fainter) cores. 
This can be quantified by the difference between the unweighted and weighted alignment parameters (\DAL):
\begin{equation} \label{eqn:DeltaAL}
	\Delta A_L = A_{L, uw} - A_{L, w}, 
\end{equation} 
and the relative value (\DALAL) is used.
A positive and large value indicates that, despite an overall elongated core distribution, massive (brighter) cores tend to be more concentrated compared to their less massive (fainter) counterparts. 
This is conceptually similar to the feature of \textit{segregation}, where different physical quantities exhibit variations in their spatial distributions among cores.

\subsection{Visual Verification of \AL} \label{sec:Exp}
To validate the ability of our alignment parameters to quantify different core distributions, we construct a series of readily parameterizable test cases. 
From Equation~\ref{eqn:ALuw}, we can identify the key factors influencing the alignment parameter: core positions and the characteristic minor axis length (used in Equation~\ref{equ:Sij}).
We then create 1000 test clumps as two-dimensional (2D) images with cores positioned on a plane to test its performance. 
For this process, we use the following methods:
\begin{description}
	\item[(a) Select the core number] The number of cores varies from 5 to 20, chosen randomly with a uniform distribution.
	\item[(b) Limit the core position] The cores are positioned on a 2D plane within an ellipse with a fixed area but variable aspect ratio ranging from 1 to 3.
	\item[(c) Assign random weights] Each core is assigned a $w_i$ value chosen uniformly and randomly between 1 and 5.
	\item[(d) Evaluate the minor axis length] Once core positions and weights are determined, $\sigma_m$ is calculated using unweighted/weighted PCA and is used to normalized the core separations in Equation~\ref{equ:Sij}.
	\item[(e) Calculate the alignment parameters] \ALuw  and \ALw are calculated using Equations~\ref{eqn:ALuw} and~\ref{eqn:ALw}, respectively.
\end{description}

\begin{figure*}[htb!]
	\epsscale{1.1}
	\plotone{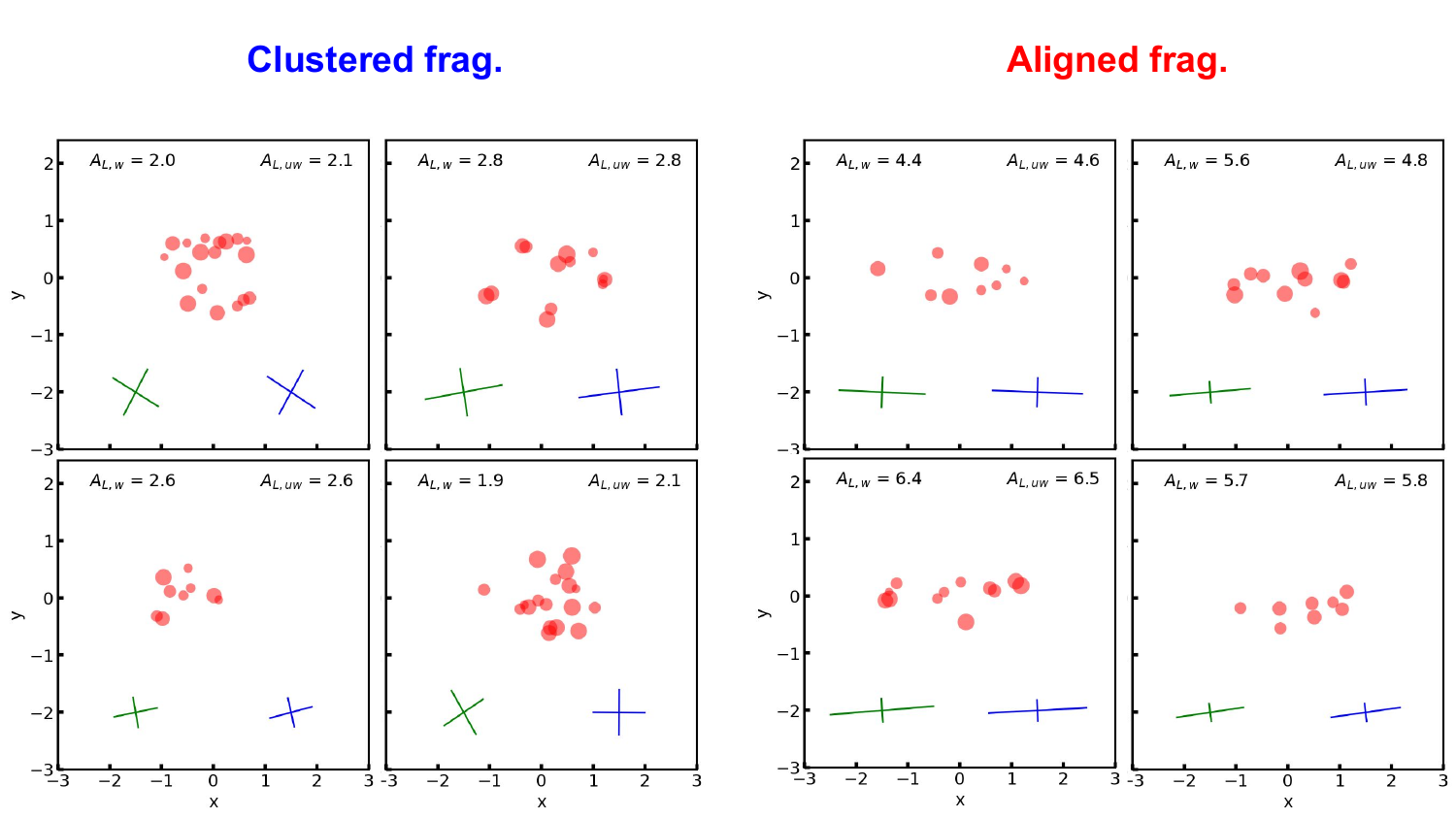}
	\caption{Examples of core distributions generated using the method described in Section~\ref{sec:Exp}. The size of each core is proportional to its assigned $w_i$. The major and minor axis lengths obtained by weighted PCA (green) and unweighted PCA (blue) are visualized in each panel. The corresponding values of \ALuw and \ALw are also displayed. The left four panels show clumps visually classified as \textquote{clustered fragmentation}, while the right four panels depict \textquote{aligned fragmentation}.}
	\label{fig:ALExpEx}
\end{figure*}

In addition to calculating the alignment parameters (\ALuw and \ALw) for each test clump, we also categorize them as either \textquote{aligned} or \textquote{clustered} based on visual inspection of the core positions in the data \footnote{To simplify the visual inspection task, a subset of test clumps is classified into either \textquote{aligned} or \textquote{clustered} by multiple individuals due to subjectivity in visual classification and is discussed in detail in Section~\ref{sec:ALBoundary}.}. 
This categorization leverages human intuition, which is well-suited for this task based solely on the relative positions of the cores. 
The categorization serves as a benchmark to compare with \ALuw values, allowing us to evaluate the effectiveness of the alignment parameters in capturing the visual assessment of core alignment.
Figure~\ref{fig:ALExpEx} displays a few examples generated using the described process. 
The left four panels show the test clumps visually classified as \textquote{clustered fragmentation}, while the right four panels depict \textquote{aligned fragmentation}. 
For each panel, the major and minor axis lengths obtained by weighted PCA (green) and unweighted PCA (blue) are visualized. 
The corresponding values of \ALuw and \ALw are also displayed. 
The size of each core is proportional to its own $w_i$.
As expected, higher values of the alignment parameter (\ALuw) are associated with cases where cores are more aligned and exhibit greater separation.

\begin{figure}[htb!]
	\epsscale{1.1}
	\plotone{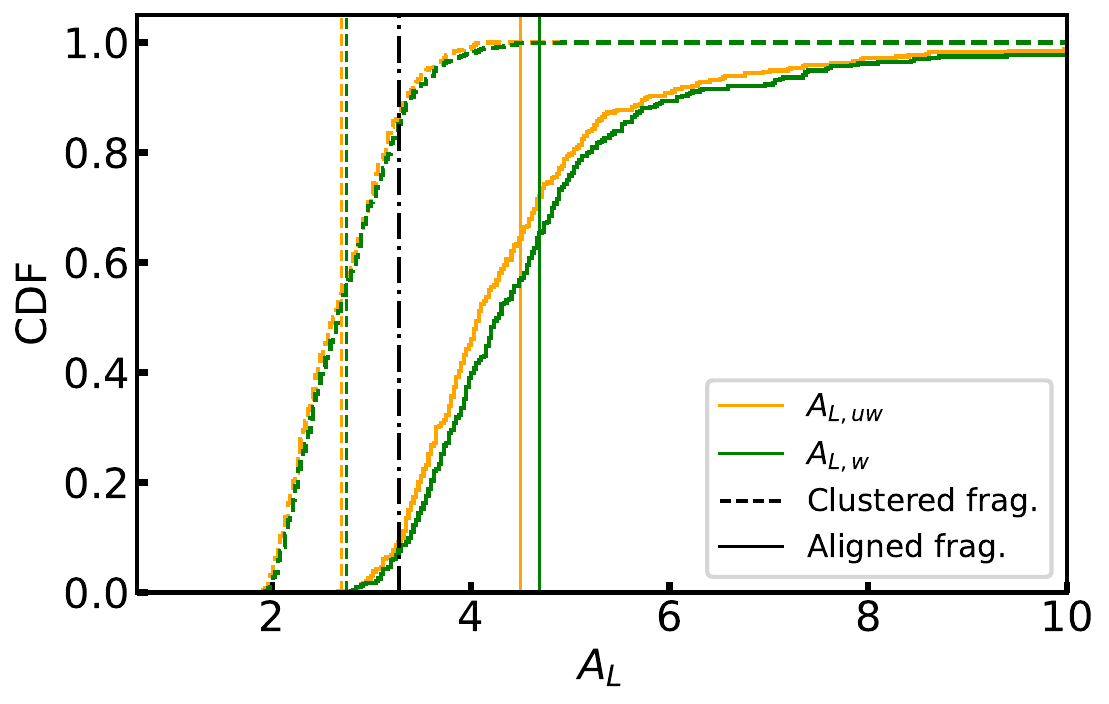}
	\caption{The CDFs of \ALuw (orange) and \ALw (green) for the two core alignment categories based on 1000 test clumps generated in Section~\ref{sec:Exp}. The dashed-line curves represent the cases values classified as \textquote{clustered fragmentation} category, while the solid-line curves correspond to the \textquote{aligned fragmentation} category. The mean values for each category and alignment parameter are indicated by the vertical lines. The black dot-dashed line indicates the threshold derived in Section~\ref{sec:ALBoundary} for the two categories. Notably, the $x$-axis is capped to highlight the distinction between the two distributions.}
	\label{fig:ALDist}
\end{figure}

Figure~\ref{fig:ALDist} presents the cumulative distribution functions (CDFs) of the alignment parameters (\ALuw as orange lines and \ALw as green lines) for the two core alignment categories based on 1000 test clumps. 
The dashed-line curves represent the distributions for cases classified as \textquote{clustered fragmentation}, while the solid-line curves correspond to \textquote{aligned fragmentation} .
The figure reveals a clear distinction between the distributions of the two categories. 
Notably, 607 test clumps were classified as \textquote{clustered fragmentation} with a mean \ALuw of 2.7 and a mean \ALw of 2.8. 
In contrast, 393 clumps belonged to the \textquote{aligned fragmentation} category with a mean \ALuw of 4.5 and a mean \ALw of 4.7. 
This clear separation supports the potential of our alignment parameters to differentiate between these core distribution types.

Furthermore, the CDFs for \textquote{aligned fragmentation} exhibits a long tail towards higher values compared to \textquote{clustered fragmentation} (the $x$-axis in Figure~\ref{fig:ALDist} is capped to emphasize this difference in low value). 
This behavior aligns with our expectations. 
After all, the alignment parameters reflect the average core separation within a clump. 
Larger separations naturally lead to higher values, which can be interpreted as a signature of \textquote{aligned fragmentation}.

\begin{figure*}[htb!]
	\plotone{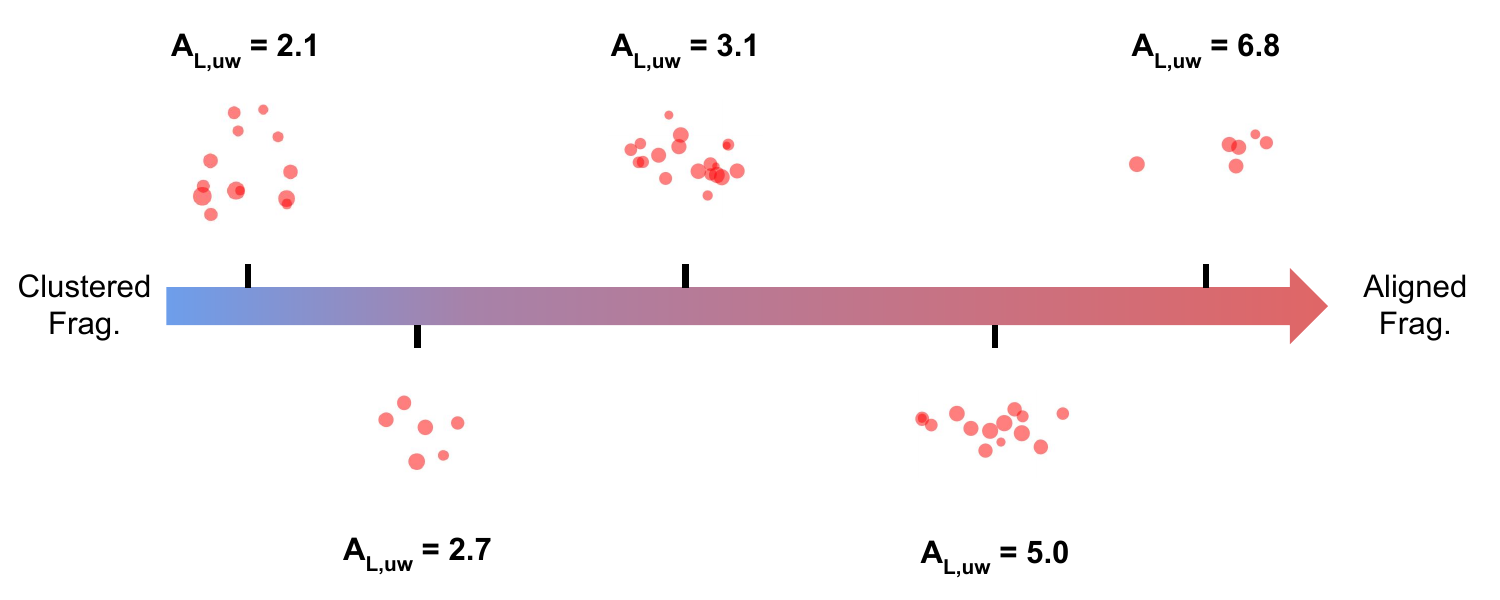}
	\caption{Examples of core distributions generated using the method described in Section~\ref{sec:Exp}. The size of each core is proportional to its $w_i$. The corresponding value of \ALuw is displayed above or below each image. These panels are arranged such that \ALuw increases from left to right, with higher values visually resembling \textquote{aligned fragmentation}.}
	\label{fig:ALuwIllus}
\end{figure*}

In summary, our alignment parameters offer a key advantage: they provide an automated, reproducible, and quantitative method for measuring core alignment, eliminating the subjectivity inherent in visual inspection.
This sensitivity to core distribution allows the parameters to effectively capture the differences previously only identified through visual inspection. 
Figure~\ref{fig:ALuwIllus} illustrates the overall trend of \ALuw with respect to different core distributions.

\subsection{An \AL Threshold for Classification} \label{sec:ALBoundary}
For practical applications, establishing a threshold value for \ALuw could be beneficial as a two-label classification tool. 
With such a threshold, subsequent analysis could then focus on identifying potential difference in physical properties between these two groups.

We now describe the procedure for obtaining a threshold value for \ALuw. 
\begin{description}
	\item[(a) Balanced Sample Size] Since our test clumps in Section~\ref{sec:Exp} are not evenly distributed between the \textquote{aligned} and \textquote{clustered} categories (with more clustered clumps), we first balance the sample size. This is achieved by randomly selecting a subset of clustered clumps such that the number of samples in each category becomes equal.
	\item[(b) Label Assignment] We assign labels based on visual identification as the \textit{true} label ($P_{true}$): 0 for \textquote{clustered} and 1 for \textquote{aligned} clumps. Then, for each test clump and a candidate threshold value ($\gamma$), a \textit{predicted} label ($P_{pred}$) is obtained using the following equation:
	\begin{equation}
	P_{pred} (\gamma) =
	\begin{cases}
		1, &\text{if } A_{L, uw} > \gamma \\
		0, &\text{if } A_{L, uw} \leq \gamma
	\end{cases}.
	\end{equation}
	\item[(c) Threshold Selection] A loss function ($\mathcal{L}$) is defined to quantify the misclassification:
	\begin{equation} \label{eqn:Loss}
	\mathcal{L} (\gamma) = \frac{1}{N_e} \sum_{i=1}^{N_e} |P_{pred, i} (\gamma) - P_{true, i}|,
	\end{equation}
	where $N_e$ is the total sample size after balancing. The optimal threshold ($\gamma_t$) minimizes the loss function:
	\begin{equation}
	\frac{d \mathcal{L} }{d \gamma}\vert_{\gamma = \gamma_t} = 0.
	\end{equation}
	\item[(d) Bootstrapping for Robustness] To improve the robustness of the threshold estimation, Step (a) to (c) are repeated 500 times. The final threshold value ($\gamma_{t, avg}$) is then determined as the average of the obtained optimal thresholds ($\gamma_t$) across all bootstrap runs.
\end{description}

\begin{figure}[htb!]
	\epsscale{1.1}
	\plotone{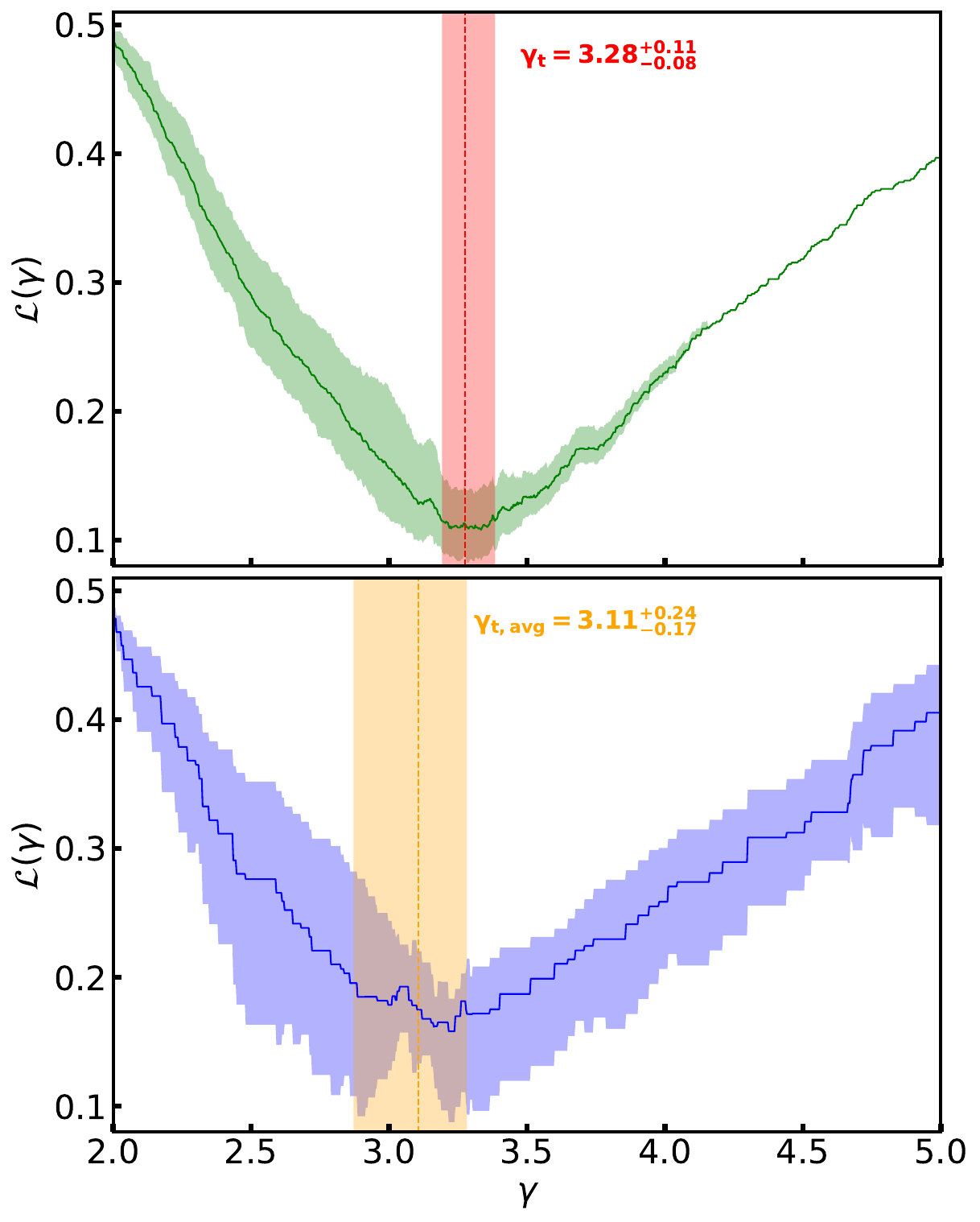}
	\caption{Loss function ($\mathcal{L}$ ) as a function of the candidate threshold value ($\gamma$). The upper panel shows the results obtained for a balanced sample size of $N_e = 2 \times 393 = 786$ (corresponding to the sub-sample of the 1000 samples from Section~\ref{sec:Exp}). The lower panel presents the variability in the mean loss function and $\gamma_{t, avg}$ across classifications of the same 100 test clumps by 6 people. The shaded regions all indicate the range of the values.}
	\label{fig:ALuwThres}
\end{figure}

Figure~\ref{fig:ALuwThres} depicts $\mathcal{L}$ as a function of $\gamma$. 
The upper panel shows the relationship for a balanced sample size of $N_e = 2 \times 393 = 786$, taken from the total sample of 1000 realizations discussed in Section~\ref{sec:Exp}.
To assess the uncertainty in the threshold determination due to labeling by a single individual, a sub-sample of 100 test clumps from Section~\ref{sec:Exp} was evaluated by 6 different people in total for classification.
Each set of classifications was then subjected to the bootstrapping process to obtain its own $\gamma_{t, avg}$.
The lower panel of Figure~\ref{fig:ALuwThres} illustrates the variability in the averaged loss function and the range of $\gamma_{t, avg}$ across classifications by different people.
Notably, the upper panel of Figure~\ref{fig:ALuwThres} suggests $\gamma_{t, avg} \sim 3.3$. 
This value is close to the average obtained across classifications by different people ($\text{mean}(\gamma_{t, avg}) \sim 3.1$), as shown in the lower panel.
Consequently, a threshold value of \ALuw = 3.3 can be used to classify core arrangements as either \textquote{clustered} or \textquote{aligned}.
This threshold is visualized as a black dot-dashed line in Figure~\ref{fig:ALDist}. 

Using this threshold, the misclassification rates for the 1000 test clumps in Section~\ref{sec:Exp} are $14\%$ for the \textquote{clustered} category (i.e., clumps labeled clustered but classified as aligned by the threshold) and $9\%$ for the \textquote{aligned} category.
The observed asymmetry in the misclassification rates between the two categories can be attributed to the larger number of samples in the \textquote{clustered} category.

\subsection{Caveats} \label{sec:caveat}
Equations~\ref{eqn:ALuw} and~\ref{eqn:ALw} highlight that core properties are crucial for calculating alignment parameters, thus influencing their interpretation. 
However, limitations arise from solely using 2D core distributions. 
Since observations are typically made on the plane-of-the-sky (POS), we can't capture the true 3D core structure due to projection effects. 
This is a common challenge for all 2D-based parameters.

\section{Applying \AL to Observations} \label{sec:CorrObs}
In this section, we utilize 1.3 mm dust continuum images from the ASHES survey \citep{2019ApJ...886..102S, 2023ApJ...950..148M} to investigate potential relationships between clump properties and core alignment. 
For this purpose we employ Kendall's rank correlation, a statistical measure of association between two datasets \citep{KendallTest}. 
For ASHES's samples size, 39 data points, assuming no ties, a Kendall's $\tau$ value of 0.219 is required to achieve a 95\% confidence level, while $\tau$ values of 0.287 and 0.366 are needed for 99\% and 99.9\% confidence levels, respectively.
Thus, with this sample size, we may robustly detect any moderate or strong correlation ($\tau > 0.25$) with high confidence in its statistical significance.

A summary of the correlations, detailed in later sections, is presented in Table~\ref{tab:TauPNominal}. 
The table includes Kendall's $\tau$ values and corresponding $p$-values. Appendix~\ref{app:Uncertainty} provides a detailed discussion of the uncertainties associated with each parameter and their potential impact on the correlations presented in this work.

\begin{deluxetable}{@{\extracolsep{4pt}}lcc@{}}
	\tablecaption{Statistical results of the correlation analysis using Kendall's rank correlation though Sections~\ref{sec:CorrFrag} to~\ref{sec:Segregation}. \label{tab:TauPNominal}} 
	\tablehead{
		\colhead{Parameter} & \colhead{\ALuw} & \colhead{\ALw} 
	}
	\startdata
	$\delta_{sep, avg}/\lambda_{J, cl}^{th}$ &  0.012 (0.913) &  -0.074 (0.506)  \\
	$\delta_{sep, avg}/\lambda_{J, cl}^{tur}$ & -0.034 (0.762) & -0.179 (0.108) \\
	Core Number & -0.095 (0.403) & -0.254 (0.025) \\
	$M_{cl}$ & -0.062 (0.585) & 0.082 (0.473) \\
	$L_{cl}$ & -0.096 (0.390) & 0.007 (0.952) \\
	$R_{cl}$ & -0.020 (0.856) & 0.217 (0.054) \\
	$T_{cl}$ & -0.112 (0.321) & -0.087 (0.438) \\
	$n_{cl}$ & -0.026 (0.818) & -0.220 (0.049) \\
	$\sigma_{cl, v}$ & -0.016 (0.893) & 0.089 (0.442) \\
	$\alpha_{vir}$ & -0.005 (0.961) & 0.003 (0.981) \\
	$R_1$ & 0.107 (0.339) & 0.053 (0.637) \\
	$R_2$ & -0.023 (0.837) & -0.088 (0.432) \\
	CFE & -0.077 (0.490) & -0.255 (0.022) \\
	$f$(proto) & 0.009 (0.943) & 0.028 (0.830) \\
	$L/M$ & -0.107 (0.339) & -0.063 (0.570)  \\
	\hline \hline
	\multicolumn{1}{c}{Parameter} & \colhead{\DALAL}  \\
	\hline
	$\Lambda_{\text{MSR}}$ & 0.136 (0.222) \\
	\enddata
	\tablecomments{For each column, Kendall's $\tau$ values are displayed, with corresponding $p$-values in parentheses.}
\end{deluxetable}

\subsection{ASHES Observation} \label{sec:ASHES}
The ASHES survey targets thirty-nine 70 $\mu$m dark infrared dark clouds (IRDCs) with the potential to form high-mass stars and that are in their early stages of development.
The ASHES sample was selected to include only IRDC clumps in their early evolutionary stage \citep{2022ApJ...936...80S, 2023ApJ...949..109L, 2023ApJ...950..148M}, making them ideal to investigate how core alignment relates to young star formation environments.

The ASHES survey observations were conducted with the Atacama Large Millimeter/submillimeter Array (ALMA) in Band 6 over three observing cycles: Cycle 3 (2015.1.01539.S, PI: P. Sanhueza), Cycle 5 (2017.1.00716.S, PI: P. Sanhueza), and Cycle 6 (2018.1.00192.S, PI: P. Sanhueza). 
The IRDC clumps in the ASHES survey are all massive ($\gtrsim 500\ M_{\odot}$), dense ($\gtrsim 5 \times 10^3$ cm$^{-3}$), 70 $\mu$m dark in \textit{Herschel} survey, and within 6 kpc (see \citealp{2019ApJ...886..102S} and \citealp{2023ApJ...950..148M} for detailed sample selection)). 
The dust continuum observations reached an average RMS noise level of $\sim 0.094$ mJy beam$^{-1}$ with a beam size of $\sim 1".2$. 
In total, 839 cores were extracted and with a size of $\sim 0.01-0.1$ pc within 39 clumps \citep{2023ApJ...950..148M}.

\subsubsection{Alignment Parameters on ASHES} \label{sec:ALonASHES}
This work employs the dendrogram technique \citep{Rosolowsky2008}, which is implemented in the \texttt{astrodendro} Python package \citep{2019ascl.soft07016R}, to identify cores within the 39 clumps observed by ASHES. 
This approach aligns with the core selection criteria used in \cite{2023ApJ...950..148M, 2024ApJ...966..171M} to ensure consistency and focus on the same set of 839 cores. 

The intensity-weighted mean positions identified for the \textit{leaf} structures by the dendrogram are used as core positions. 
The integrated flux (after correcting for primary beam attenuation) within each \textit{leaf} of the dendrogram then serves as $w_i$ for the corresponding core. 
Details regarding the identified cores, including their \ALuw and \ALw values and the major and minor axis lengths for each clump (similar to Figure~\ref{fig:ALExpEx}), are provided in Appendix \ref{app:ASHESImage}.

Figure~\ref{fig:ASHESALDist} presents a comparison of CDFs for \ALuw and \ALw. 
Using a threshold of 3.3 obtained in Section~\ref{sec:ALBoundary} for \ALuw, 35 clumps were classified as \textquote{clustered} and 4 as \textquote{aligned}. 
With the same threshold, for \ALw, 29 clumps fall into the \textquote{clustered} category and 10 into the \textquote{aligned} category. 
Notably, the distribution suggests a larger number of clumps have a higher value of \ALw compared to \ALuw, with a mean value of 2.6 for \ALuw and 3.1 for \ALw. 
This implies a potential difference in the spatial distribution of massive cores (with higher integrated flux) relative to less massive cores within the clumps.

\begin{figure}[htb!]
	\epsscale{1.1}
	\plotone{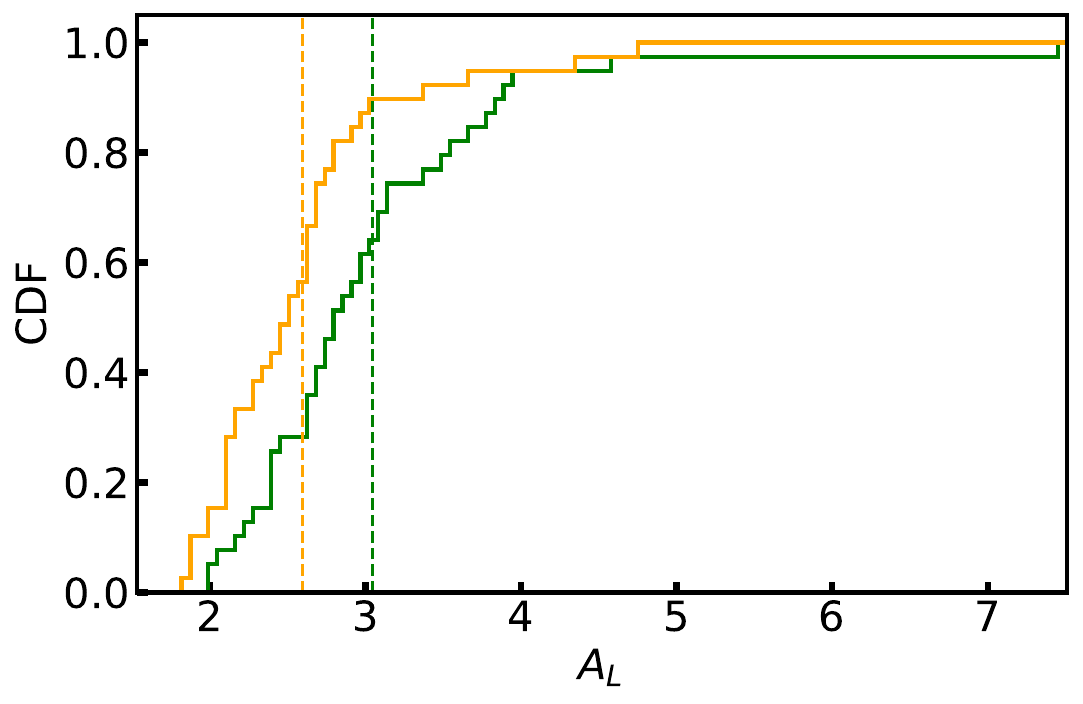}
	\caption{Comparison of CDFs for \ALuw (orange) and \ALw (green) for the ASHES clumps. The dashed lines represent the mean values of each distribution.}
	\label{fig:ASHESALDist}
\end{figure}

\subsubsection{Comparing with Low-mass Star-forming Regions}
Observations of nearby low-mass star-forming clouds (e.g. Taurus, Aquila) by Hersechel reveal that cores are often found along filaments, suggesting that these cores form via filament fragmentation (see review by \citealp{2023ASPC..534..233P}). 
One might therefore expect to find cores in a highly aligned configuration, characterized by high values of \AL. 
However, the results from the ASHES sample show a distinct preference for clustered configurations, with 35 out of 39 clumps exhibiting low \AL values.

One possible scenario is that our findings from the ASHES sample are not necessarily inconsistent with the filament fragmentation paradigm. 
The ALMA observations analyzed here have a limited field of view of $\sim$1 pc, covering only the brightest part of the clumps. 
This is significantly smaller than the typical cloud size of $>10$ pc observed in Herschel surveys. 
When considering such a small field of view within a nearby low-mass cloud, it is likely to include multiple cores along multiple filaments, which together form a complex filamentary network as observed in Herschel surveys \citep{2014prpl.conf...27A, 2015A&A...584A..91K, 2016MNRAS.459..342M}. 
Thus, it is reasonable to expect clustered alignment patterns to be common, even in regions where fragmentation occurs along filaments. 
This interpretation is supported by the high abundance of filamentary hub-like structures identified in the ASHES clumps by \cite{2023ApJ...950..148M}, with 17 out of 39 clumps exhibiting this morphology (see Section~\ref{sec:Hub} for more details).

Another possibility is that there may be fundamental differences in the fragmentation processes occurring in massive clumps, as studied in the ASHES sample, and in nearby low-mass clouds. 
Such differences could arise from the magnitude differences in physical properties, such as surface density, between these two environments. 
To definitively determine which scenario is more likely, a detailed comparison of alignment parameter distributions between ASHES clumps and cores in nearby clouds, such as those identified in Gould Belt clouds from Herschel surveys, would be necessary. 
However, this is beyond the scope of the current study.

\subsection{Measuring the Fragmentation Properties} \label{sec:CorrFrag}
Given that alignment parameters quantify core distribution within the clumps, and considering that these cores have fragmented from their parent structures, we first compare them with other parameters that characterize fragmentation properties, such as core separation and core number.

The core separation is a common measure to assess whether fragmentation is Jeans-like or not \citep{2015MNRAS.453.3785P, 2018A&A...617A.100B, 2021A&A...649A.113B, 2021ApJ...912..159P, 2023MNRAS.520.2306T, 2024A&A...682A..81B, 2024ApJ...974...95I}.
We use the core separation identified by the minimum spanning tree (MST; \citealt{MST}) method, denoted by $\delta_{sep}$. 
This value is then compared with the thermal Jeans length ($\lambda_{J, cl}^{th}$) and turbulent Jeans length ($\lambda_{J, cl}^{tur}$). 
The derivation of these Jeans lengths for the 39 clumps can be found in \cite{2024ApJ...966..171M}. 
That study suggests that thermal Jeans fragmentation, rather than turbulent fragmentation, is a dominant process in the early stages of high-mass star formation within the ASHES clumps. 
Here, we focus on comparing the average core separation ($\delta_{sep, avg}$) with alignment parameters for each clump.

\begin{figure*}[htb!]
	\epsscale{1.1}
	\plotone{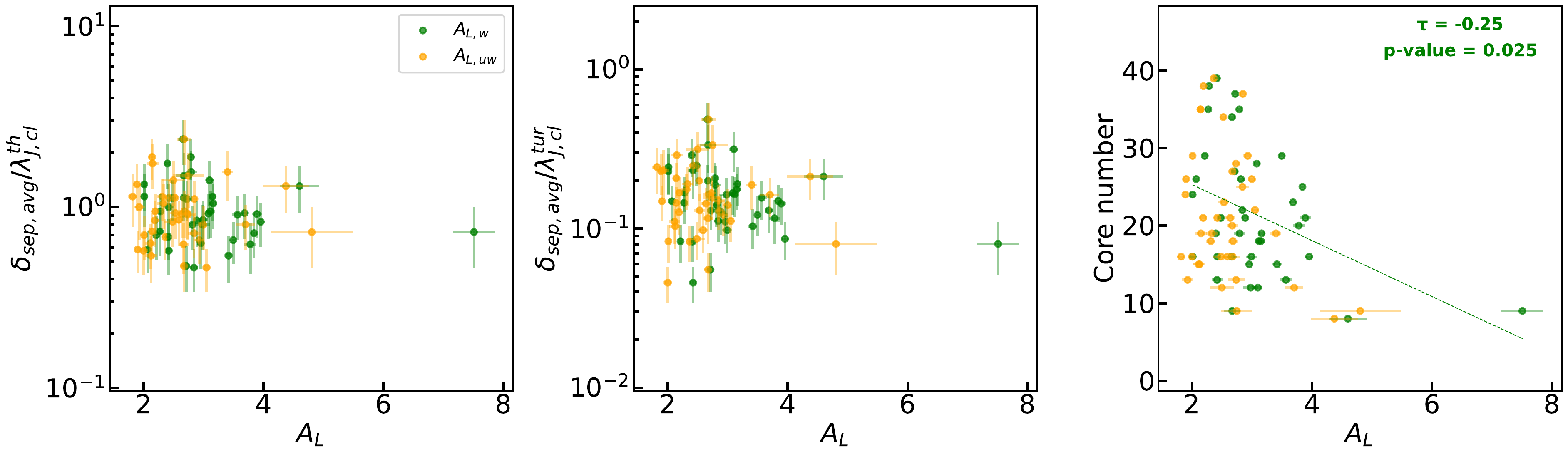}
	\caption{Relationships between \ALuw (orange) and \ALw (green) with common fragmentation parameters. The left, middle, and right panels depict the comparisons with the ratio of average core separation ($\delta_{sep, avg}$) to thermal Jeans length ($\lambda_{J, cl}^{th}$), the ratio of $\delta_{sep, avg}$ to turbulent Jeans length ($\lambda_{J, cl}^{tur}$), and the number of identified cores, respectively. The Kendall's $\tau$ and corresponding $p$-value using the nominal values are displayed for the correlations with $p$-value $< 0.05$ and a linear function is fitted to visualize the trend. The dots and error bars represent the mean value and standard deviation, respectively, as discussed in Appendix~\ref{app:Uncertainty}.}
	\label{fig:ALVSJeansLCoreNum}
\end{figure*}

The left and middle panels of Figure~\ref{fig:ALVSJeansLCoreNum} depict the relationships between alignment parameters (\ALuw, shown in orange, and \ALw, shown in green) and the ratios of $\delta_{sep, avg}/ \lambda_{J, cl}^{th}$ and $\delta_{sep, avg}/ \lambda_{J, cl}^{tur}$, respectively.
Kendall's rank correlation test is employed to quantify the association between the two sets of data. 
The resulting absolute $\tau$ values are all $<$ 0.219 and $p$-values are $>$ 0.05 (Table~\ref{tab:TauPNominal}), indicating that the correlations are weak, and we cannot reject the null hypothesis that any apparent correlation may be due solely to randomness.

We next investigate the relationship between alignment parameters and the number of identified cores, also referred to as the \textit{fragmentation level} \citep{2015MNRAS.453.3785P, 2021ApJ...912..159P}. 
All cores within the clumps are included since they contribute to the calculation of the alignment parameters. 
Previous studies by \cite{2021ApJ...912..159P} and \cite{2024ApJ...966..171M} demonstrated a positive correlation between core number and surface density, suggesting that denser environments tend to harbor more cores, aligning with expectations from thermal Jeans fragmentation.
The right panel of Figure~\ref{fig:ALVSJeansLCoreNum} presents the relationships between alignment parameters and the number of identified cores. 
Interestingly, \ALw exhibits a weakly negative correlation with the number of cores ($\tau = -0.25$ and $p$-value = 0.025), whereas \ALuw shows no significant trend. 
Statistically, with fewer cores, there is a higher likelihood of finding them aligned. 
This feature is captured by the alignment parameters. 
Using the test clumps from Section~\ref{sec:Exp}, we consistently observe negative ($\tau < -0.20$) and significant ($p$-values $\sim 0$) correlations between the two alignment parameters and the inputted core number.
Therefore, the correlation found here is not surprising.

Overall, considering fragmentation mechanisms, these correlations suggest that alignment parameters primarily focus on quantifying the degree of core alignment, and higher or lower values do not correlate with whether the clump is undergoing thermal or turbulent fragmentation.

\subsection{Clump Properties} \label{sec:ClumpProp}
We next investigate potential correlations between alignment parameters and various properties of their host clumps in the ASHES survey. 
These properties encompass:
\begin{description}
	\item[Basic properties] clump mass ($M_{cl}$), luminosity ($L_{cl}$), radius ($R_{cl}$), dust temperature ($T_{cl}$), and averaged number density ($n_{cl}$).
	\item[Stability parameters] C$^{18}$O ($J = 2-1$) velocity dispersion ($\sigma_{cl, v}$) and virial parameter ($\alpha_{vir}$).
	\item[Morphology] the values of $R_1$ and $R_2$.
	\item[Additional parameters] core formation efficiency (CFE) and the fraction of protostellar cores to all bound cores in each clump ($f$(proto)).
\end{description}

The virial parameter, defined as $\alpha_{vir} = 5 \sigma_{cl, v}^2 R_{cl}/G M_{cl}$ (assuming a spherical clump with uniform density), quantifies the balance between the clump's gravitational energy and its kinetic energy \citep{1992ApJ...395..140B}.
The clump luminosity is calculated using Equation 3 from \cite{2017MNRAS.466..340C} (as employed in \citealp{2022ApJ...936...80S}). 
The values of $R_1$ and $R_2$ quantify elongation and central condensation within the clumps, respectively. 
A higher $R_1$ indicates a more elongated structure, while a larger $R_2$ suggests a more centrally condensed structure.
The values for $R_1$ and $R_2$ are calculated using the Python library \texttt{RJ-plots} (see \citealp{2022MNRAS.516.2782C} for details).
The remaining parameters are obtained from the ASHES survey, although only 30 clumps have available values for $f$(proto) \citep{2023ApJ...950..148M, 2024ApJ...966..171M}.

\begin{figure*}[htb!]
	\epsscale{1.1}
	\plotone{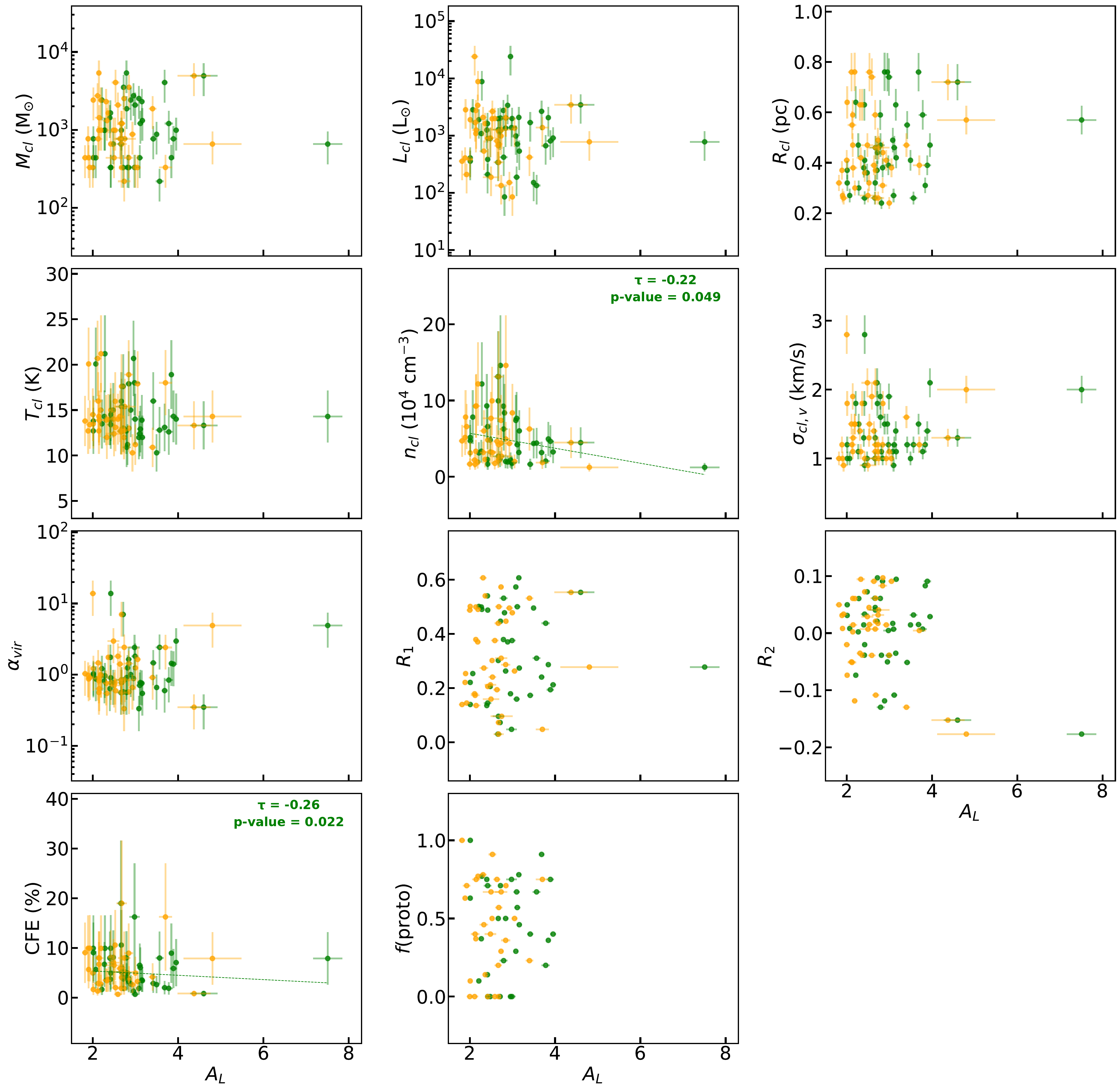}
	\caption{Relationships between \ALuw (orange) and \ALw (green) with various properties of their host clumps in ASHES. The configuration of the figures is the same as Figure~\ref{fig:ALVSJeansLCoreNum}.}
	\label{fig:ALVSProperties}
\end{figure*}

Figure~\ref{fig:ALVSProperties} presents the detailed relationships between these parameters and alignment parameters. 
The value of Kendall's $\tau$ is displayed for correlations with $p$-values $<$ 0.05.
Additionally, a linear fit is included to visualize the trend for those significant correlations.

Interestingly, our analysis reveals potential correlations only for \ALw. 
Among the clump properties, only $n_{cl}$ and CFE exhibit statistically significant but weak correlations with \ALw ($p$-value $<$ 0.05 and $\tau \sim 0.2$). 
It's important to note that $n_{cl}$ is derived from $M_{cl}$ and $R_{cl}$, so this correlation might not be entirely independent.
The observed correlations solely with \ALw suggest a possibly stronger connection between more massive cores and the host clump's properties.
Notably, the negative correlation between \ALw and CFE is intriguing. 
CFE, representing the ratio of total core mass to clump mass, implies that clustered fragmentation (low \ALw) might be a more efficient process in transferring mass from the clump scale to the core scale compared to aligned fragmentation (high \ALw). 
However, further investigation is warranted due to the weak correlations and limited number of clumps in the high \ALw regime.

To assess the influence of potential outliers, we specifically examined the relationship without the clump (G033.33) exhibiting a very high \ALw value (= 7.5). 
After excluding G033.33 and recalculating the correlations, only the one with CFE ($\tau$ = -0.30, $p$-value = 0.009) remained statistically significant. 
This strengthens the evidence for a likely correlation between \ALw and CFE.
Beside that, other correlations appear weak and statistically insignificant.

\subsection{Clump Evolution} \label{sec:ClumpEvo}
Several studies suggest that cores within star-forming clumps can interact through gravitational forces via two-body relaxation or by sinking towards the central potential well after fragmenting from their parent structures as the clump evolves. 
These interactions are expected to influence the separation between cores \citep{2018A&A...617A.100B, 2021A&A...649A.113B, 2023MNRAS.520.2306T, 2024ApJS..270....9X, 2024ApJ...974...95I}. 
Consequently, it is of interest to investigate how alignment parameters, which quantify the degree of core alignment based on a single snapshot in time, reflect potential changes in core alignment during clump evolution.

\begin{figure}[htb!]
	\epsscale{1.1}
	\plotone{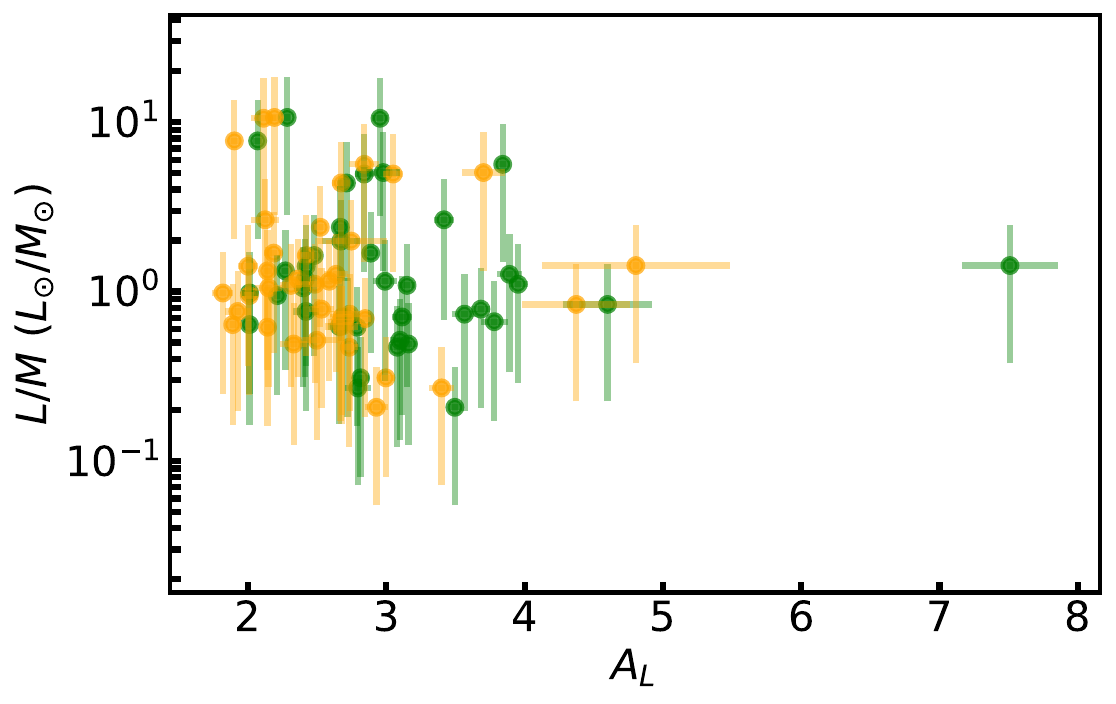}
	\caption{Relationship between \ALuw (orange) and \ALw (green) with the luminosity-to-mass ratio ($L/M$) for the ASHES clumps. The configuration of the figures is the same as Figure~\ref{fig:ALVSJeansLCoreNum}.}
	\label{fig:ALVSLM}
\end{figure}

A common tracer of clump evolution is the luminosity-to-mass ratio ($L/M$). 
As a clump progresses beyond its prestellar stage, the formation of a central young star increases its bolometric luminosity for a given mass, leading to an expected rise in $L/M$ \citep{2008A&A...481..345M, 2019MNRAS.486.4508M}. 
Figure~\ref{fig:ALVSLM} explores the relationship between alignment parameters and $L/M$ for the ASHES clumps. 
In both cases, the Kendall's tau test returns absolute $\tau < 0.219$ and $p$-values much greater than 0.05. 
Thus, any apparent weak correlation could result from randomness.

The absence of a significant correlation between alignment parameters and $L/M$ might suggest a lack of an evolutionary trend for core alignment. 
One possible explanation could be the presence of significant variations in the initial core alignment among different clumps, which could obscure any potential relationship with evolutionary stage.
However, it is also important to note that the $L/M$ values within the ASHES sample exhibit a relatively narrow range. 
Future analyses utilizing datasets encompassing a wider range of evolutionary stages would be more suitable for definitively assessing the relationship between core alignment and clump evolution. 

\subsection{Segregation} \label{sec:Segregation}
Section \ref{sec:DevelopeAL} introduces \DALAL as a parameter sensitive to the spatial distribution of cores with varying masses (fluxes).  
We also propose its potential application in investigating core segregation, where massive cores tend to be more spatially concentrated compared to lower-mass cores.
To assess the effectiveness of \DALAL  in tracing core segregation, we perform a simple correlation analysis between \DALAL and the mass segregation ratio ($\Lambda_{\text{MSR}}$, Equation 1 in \citealp{2009MNRAS.395.1449A}) derived and used by the ASHES survey \citep{2019ApJ...886..102S, 2023ApJ...950..148M}.
Since $\Lambda_{\text{MSR}}$ is a function of the number of most massive cores ($N_{\text{MSR}}$), the reported $\Lambda_{\text{MSR}}$ values here represent the maximum values obtained from the function for each clump when $N_{\text{MSR}} > 3$.

\begin{figure}[htb!]
	\epsscale{1.1}
	\plotone{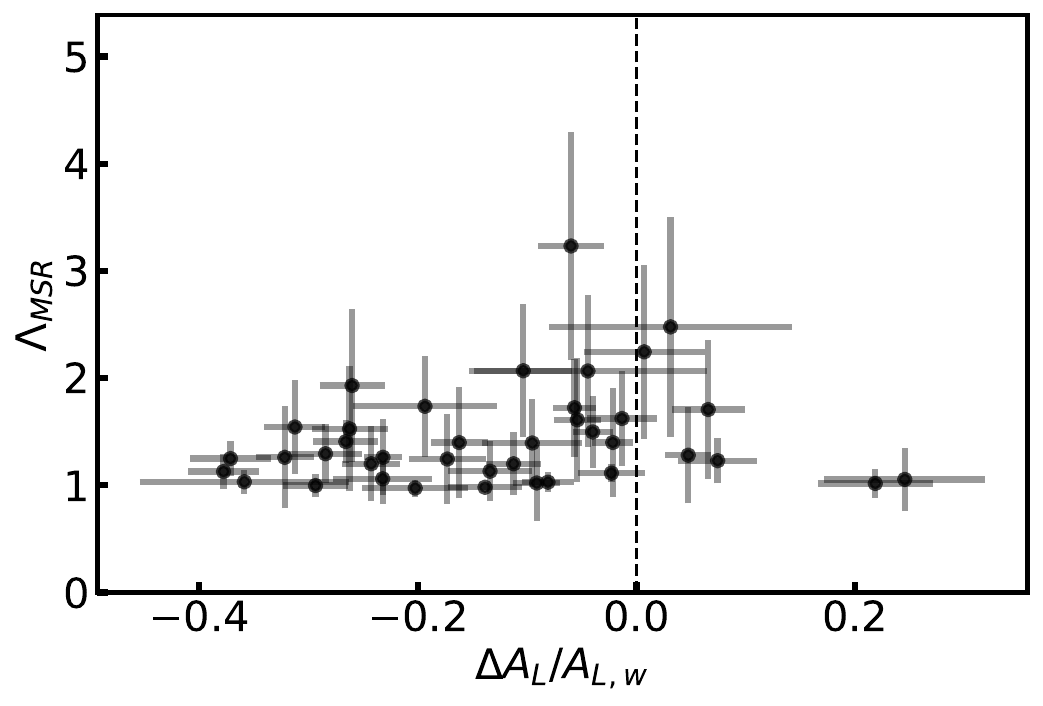}
	\caption{Relationship between \DALAL and mass segregation ratio ($\Lambda_{\text{MSR}}$) for the ASHES clumps.}
	\label{fig:DALSegregation}
\end{figure}

Figure~\ref{fig:DALSegregation} compares \DALAL and $\Lambda_{\text{MSR}}$. 
The results of Kendall's rank correlation test ($\tau$ = 0.14 and $p$-value = 0.22) indicate no statistically significant correlation. 
This might not be surprising, as caution is necessary when comparing segregation metrics due to their potentially different definitions of \textquote{segregation} \citep{2015MNRAS.449.3381P}.
For instance, the $\Lambda_{\text{MSR}}$ metric assesses whether the most massive cores are positioned closer together relative to a random distribution \citep{2009MNRAS.395.1449A, 2015MNRAS.449.3381P}. 
In contrast, \DALAL metric focuses on the relative alignment of massive cores compared to less massive ones.

While the correlation analysis may not be conclusive, a visual inspection of the clump images in Appendix \ref{app:ASHESImage} provides some insights. 
The first two clumps with the highest \DALAL values (0.24 for G022.69 and 0.22 for G028.27) do exhibit signs of overall core elongation, but with massive cores (represented by larger dot sizes) appearing more clustered. 
Conversely, the three clumps with the lowest \DALAL values (G034.13: -0.38, G340.23: -0.37, and G033.33: -0.36) show that massive cores are either more separated or the existence of a distant subcluster.
Overall, using \DALAL, we find that 32 clumps ($82\%$) have negative values, suggesting that massive cores tend to be more aligned than less massive cores in the ASHES sample.  
Furthermore, no significant correlations were found between \DALAL and any of the clump properties examined in Sections~\ref{sec:CorrFrag} to~\ref{sec:ClumpEvo}. 
All absolute values of Kendall's $\tau$ and $p$-values were smaller than 0.219 and larger than 0.05, respectively.

\subsection{Clumps with Hubs} \label{sec:Hub}
Instead of directly forming from the host clump, the cores can fragment from intermediate-scale structures, such as sub-filaments. Once sub-filaments converge to form hubs, these less symmetrical structures can provide an alternative pathway to produce clustered core distributions. These hubs are commonly observed and are associated with high-mass star formation \citep{2009ApJ...700.1609M, 2014prpl.conf...27A, 2014ApJ...791..124G, 2024MNRAS.528.1460R, 2018A&A...613A..11W, 2020A&A...642A..87K, 2024MNRAS.527.4244S}. In the ASHES sample, \cite{2023ApJ...950..148M} identified sub-filaments from the 1.3 mm dust continuum maps and visually identified hub-like features in 17 clumps, allowing us to test whether there are differences in core distribution and clump properties between these hub and non-hub clumps.

\begin{figure}[htb!]
	\epsscale{1.1}
	\plotone{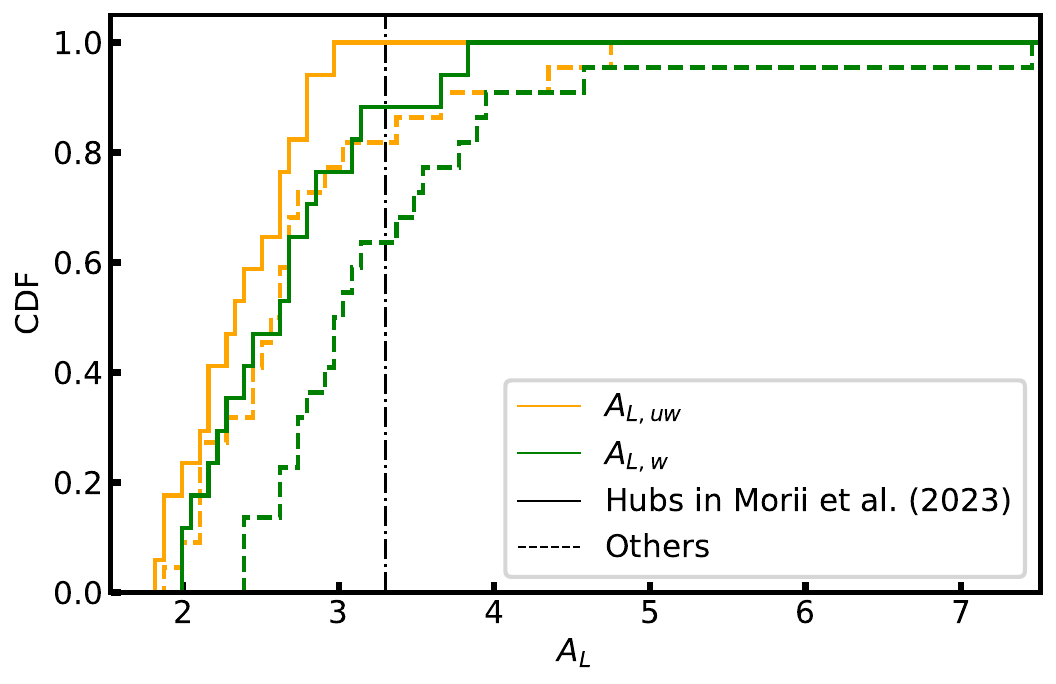}
	\caption{Comparison of CDFs of \ALuw (orange) and \ALw (green) for the hub clumps classified by \cite{2023ApJ...950..148M}  (solid lines) and the remaining clumps (dashed lines). The black dot-dashed line indicates \AL = 3.3.}
	\label{fig:ALbyHub}
\end{figure}

Figure~\ref{fig:ALbyHub} presents the comparison of \ALuw (orange) and \ALw (green) distributions for clumps classified as hubs by \cite{2023ApJ...950..148M} (solid lines) and the remaining clumps (dashed line). 
The figure clearly indicates a tendency for hub clumps to exhibit lower alignment parameter values, suggesting that hubs are more associated with clustered core distributions. 
This is expected, as alignment parameters solely consider core positions and weights, regardless of whether the cores are embedded in other substructures.

We additionally test whether there are differences between hub clumps and non-hub clumps for the clump properties explored in Sections~\ref{sec:CorrFrag} to \ref{sec:Segregation}. 
Two-sample Kolmogorov-Smirnov (KS) test was used to determine if two distributions are likely the same \citep{1958ArM.....3..469H}. 
Overall, there are no differences between the two, but only the number of identified cores (KS statistic = 0.49, $p$-value = 0.01) and \DALAL (KS statistic = 0.43, $p$-value = 0.04) exhibit significant differences.
For them, we can reject the hypothesis that the two samples come from the same distribution.
This shows that clumps forming hubs tend to have more cores, which suggests that hubs promote core formation. 
Additionally, their \DALAL distribution is narrower and more centered around zero than that of the remaining clumps (not shown), suggesting that both massive and less massive cores exhibit clustered distributions within these hub regions.

Finally, for the correlations between alignment parameters (\ALuw and \ALw) and clump-scale properties (i.e., the parameters in Sections~\ref{sec:ClumpProp} and \ref{sec:ClumpEvo}) for these two groups, we still do not observe any strong correlations. 
This suggests that even though cores may form through different fragmentation processes (e.g., in hub or non-hub environments) and exhibit differences in their distributions, there are no significant differences in the clump properties that would impact our findings.

\subsection{Absence of Strong Correlation} \label{sec:WeakCorrelation}
The lack of strong correlations between the alignment parameters and various clump properties suggests that the clump properties considered in this work may not be strong predictors of core alignment. 
Here, we discuss potential reasons for this result.

One possibility is that the cores within the ASHES sample formed through a common fragmentation process. 
\cite{2024ApJ...966..171M} compared core separation with both thermal and turbulent Jeans lengths, finding stronger consistency with thermal Jeans fragmentation. 
This aligns with other studies of IRDCs, which have also found that thermal fragmentation is dominant in the early stages of evolution \citep{2015A&A...581A.119B, 2018A&A...617A.100B, 2021A&A...649A.113B, 2018ApJ...855...24P, 2019ApJ...871..185L, 2020ApJ...894L..14L}. 
If this scenario holds true, the spread in \AL and the lack of a strong correlation may be a natural consequence of this single, common fragmentation process, with variations in \AL being attributed to random fluctuations.

Additionally, the lack of a clear correlation might suggest that the fragmentation process is chaotic in the mathematical sense, meaning that random fluctuations between statistically similar initial conditions may result in dissimilar fragmentation outcomes.
Turbulence, in its natural state, could be a potential source of this chaotic behavior.
It has been demonstrated by \cite{2022MNRAS.511.2702J} that by implementing different turbulence realizations in statistically identical initial conditions, their results showed a $\sim 50\%$ variation in the final star number, highlighting the chaotic nature of the process.

Finally, it is possible that we are missing the key clump properties.
From observations, we have known that the fragmentation process is influenced by competing mechanisms such as thermal and non-thermal motions, magnetic fields, and gravity \citep{2023MNRAS.520.2306T, 2018A&A...617A.100B, 2021A&A...649A.113B, 2024A&A...682A..81B, 2021ApJ...912..159P, 2019ApJ...878...10T, 2023ApJ...951...68C, 2022AJ....164..175C}. 
In environments with strong magnetic fields, the collapsing cloud tends to be flattened or filamentary, and we might expect to observe highly aligned cores \citep{2019ApJ...878...10T, 2023ApJ...951...68C, 2022AJ....164..175C, 2024ApJ...963..126G, 2021ApJ...912L..27E}. 
Additionally, the larger-scale structure in which the clump is embedded can influence the fragmentation process, such as in the case of multiple converging filaments (\citealp{2014ApJ...791..124G, 2024MNRAS.528.1460R, 2018A&A...613A..11W, 2020A&A...642A..87K}, and review by \citealp{2023ASPC..534..233P}). 
Therefore, detailed studies of magnetic fields and clump formation environments are crucial.

Among these factors, due to the turbulent nature of the molecular clouds \citep{1981MNRAS.194..809L}, we favor a chaotic fragmentation scenario to explain the weak correlations observed in the ASHES sample.
More observations, including magnetic field measurements and observations of clumps across a wider range of evolutionary stages, will be useful to investigate the mechanisms of clump fragmentation.

\subsection{Caveats for Correlations}
The correlations conducted in this section assume that the measured clump properties can accurately represent the initial conditions under a simple fragmentation model (e.g., Jeans fragmentation) and that the present-day core distribution is a direct outcome of these initial conditions. 
However, this assumption is idealized and does not account for the evolution of clumps and cores over time, or the potential influence of external factors such as stellar feedback from nearby stars. 
Even with detailed studies of gas dynamics within these regions, the chaotic nature of fragmentation could still obscure the link between initial conditions and the present-day core distribution. 
Therefore, a larger sample, derived from either observations or simulations, would be beneficial in addressing this issue and providing insights into the scatter in the relationship between clump properties and core distribution.


Additionally, when using the alignment parameters, it is crucial to verify that the identified cores are dynamically consistent with their host clump. 
For example, one can check whether the core's LSR velocity aligns with that of the clump to exclude potential foreground or background structures.
In terms of ASHES core sample, for the cores with available and sufficient signal-to-noise ratio measurements of N$_2$D$^+$ or DCO$^+$, \cite{2023ApJ...949..109L} demonstrated that their LSR velocities are consistent with that of the host clump (a full table of core LSR velocity values can be found in \citealp{2024ApJ...966..171M}).

%
%

\section{Conclusion} \label{sec:Conclusion}
Increasing availability of high-resolution and high-sensitivity data toward star-forming regions necessitates automated and reproducible methods for quantifying dense core alignment. 
This paper introduces \textquote{unweighted and weighted alignment parameters} to address this need. 
To assess the robustness of these parameters, we generate artificial test clumps and compare the results with human visual inspection. 
Additionally, we calculate \ALuw and \ALw for 39 clumps in the ASHES survey to explore potential correlations with physical parameters. 
The key findings of this study are summarized as follows:
\begin{enumerate}
	\item In Section~\ref{sec:DevelopeAL}, we introduce the unweighted alignment parameter (\ALuw) as the mean of normalized core separations. By incorporating core weights (e.g. core fluxes or masses), we derive the weighted alignment parameter (\ALw).
	\item  By comparing \ALw values for test clumps with human-assigned labels in Sections~\ref{sec:Exp} and~\ref{sec:ALBoundary}, we demonstrate that a higher value indicates a more \textquote{aligned} core arrangement, while a smaller value suggests a more \textquote{clustered} case. For two-label classification, a robust threshold of 3.3 is proposed to differentiate between these two groups.
	\item Applying the alignment parameters to the 39 ASHES clumps in Section~\ref{sec:ALonASHES} reveals a prevalence of \textquote{clustered} fragmentation. The difference between \ALuw and \ALw distributions indicates potential differences in the spatial distribution of massive and less massive cores within clumps, which are further explored using \DALAL in Section~\ref{sec:Segregation}.	
	\item We found no strong correlations between the alignment parameters and the clump properties from the ASHES sample (Sections~\ref{sec:ClumpProp} and~\ref{sec:ClumpEvo}). 
	This suggests that the fragmentation in the clump scale might  be not purely determined by the physical conditions of that scale.
	\item In Section~\ref{sec:WeakCorrelation}, we postulate several possibilities for the weak correlations: the clumps may be fragmenting in a similar manner, the fragmentation process may be inherently chaotic, or key clump properties may be missing. Considering the turbulent nature of the clumps, a chaotic fragmentation process seems more plausible. However, more observations, including magnetic field measurements and clumps across a wider range of evolutionary stages, will be necessary to determine the underlying mechanisms.
\end{enumerate}

We provide an open-source code for calculating the alignment parameters, along with an example script to generate the test clumps presented in Section~\ref{sec:Exp}.
This code is available at \cite{ALPara}.

\begin{acknowledgments}
	We thank the anonymous referee for prompt reviews and constructive suggestions to improve this paper.
	We also thank Patrick Koch, Hsi-Wei Yen, Jia-Wei Wang, and the other group members for their constructive discussions during our group meetings, which significantly contributed to the improvement of this work. 
	Additionally, we acknowledge Shang-Jing Lin, Jo-Shui Kao, Po-Hsun Lai, and Ingrid Tseng for providing visual inspection labels for clump classification, and Kaho Morii for consulting information about the ASHES survey.
	PS was partially supported by a Grant-in-Aid for Scientific Research (KAKENHI Number JP22H01271 and JP23H01221) of JSPS. 
	PS was supported by Yoshinori Ohsumi Fund (Yoshinori Ohsumi Award for Fundamental Research).
	This manuscript is supported by National Science and Techonoly Council (NSTC) through grants 112-2112-M-001-066- and 111-2112-M-001-064-.
\end{acknowledgments}


\software{NumPy \citep{harris2020array},
	            SciPy \citep{2020SciPy-NMeth},
	            Matplotlib \citep{Hunter:2007},
				Astropy \citep{astropy:2013, astropy:2018, astropy:2022}, 
				astrodendro \citep{2019ascl.soft07016R}, 
				R-J plots \citep{2022MNRAS.516.2782C}, 
				Alignment Parameters \citep{ALPara}.}
				
\appendix
\section{ASHES Images}\label{app:ASHESImage}
This section presents the 1.3 mm dust continuum images from the ASHES survey, along with the identified cores and corresponding alignment parameter values (see Section~\ref{sec:ALonASHES} for further details).

\begin{figure*}[htb!]
	\epsscale{1.1}
	\plotone{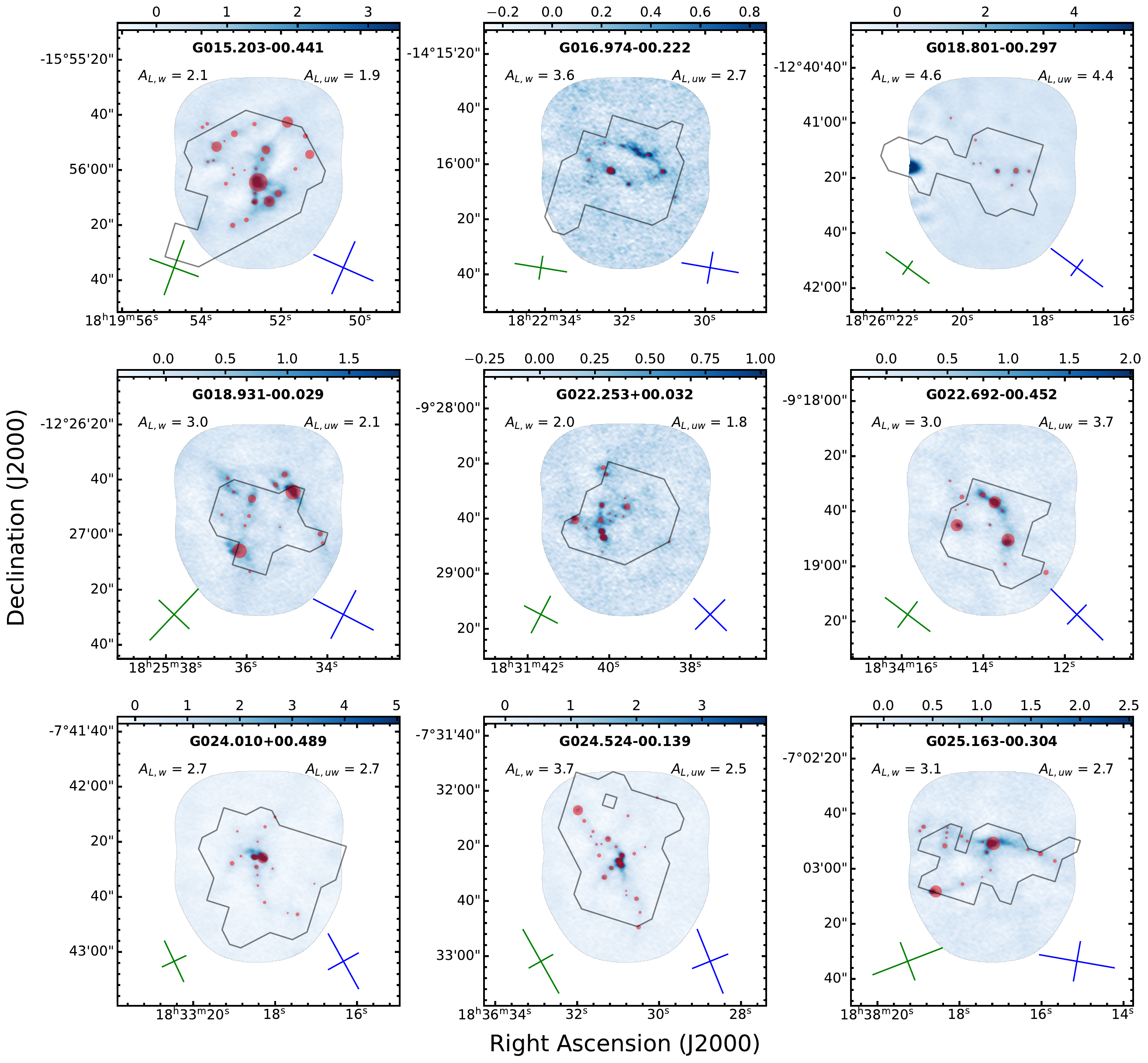}
	\caption{Each panel displays the 1.3 mm dust continuum image from the ASHES survey as the background, with colorbars in units of mJy beam$^{-1}$. Overlaid on this are grey contours delineating regions identified from the 870 $\mu$m ATLASGAL survey, used to calculate $R_1$ and $R_2$ as described in Section~\ref{sec:ClumpProp}. Other image configurations match those presented in Figure~\ref{fig:ALExpEx}. }
	\label{fig:ASHESAL_0}
\end{figure*}

\begin{figure*}[htb!]
	\epsscale{1.1}
	\plotone{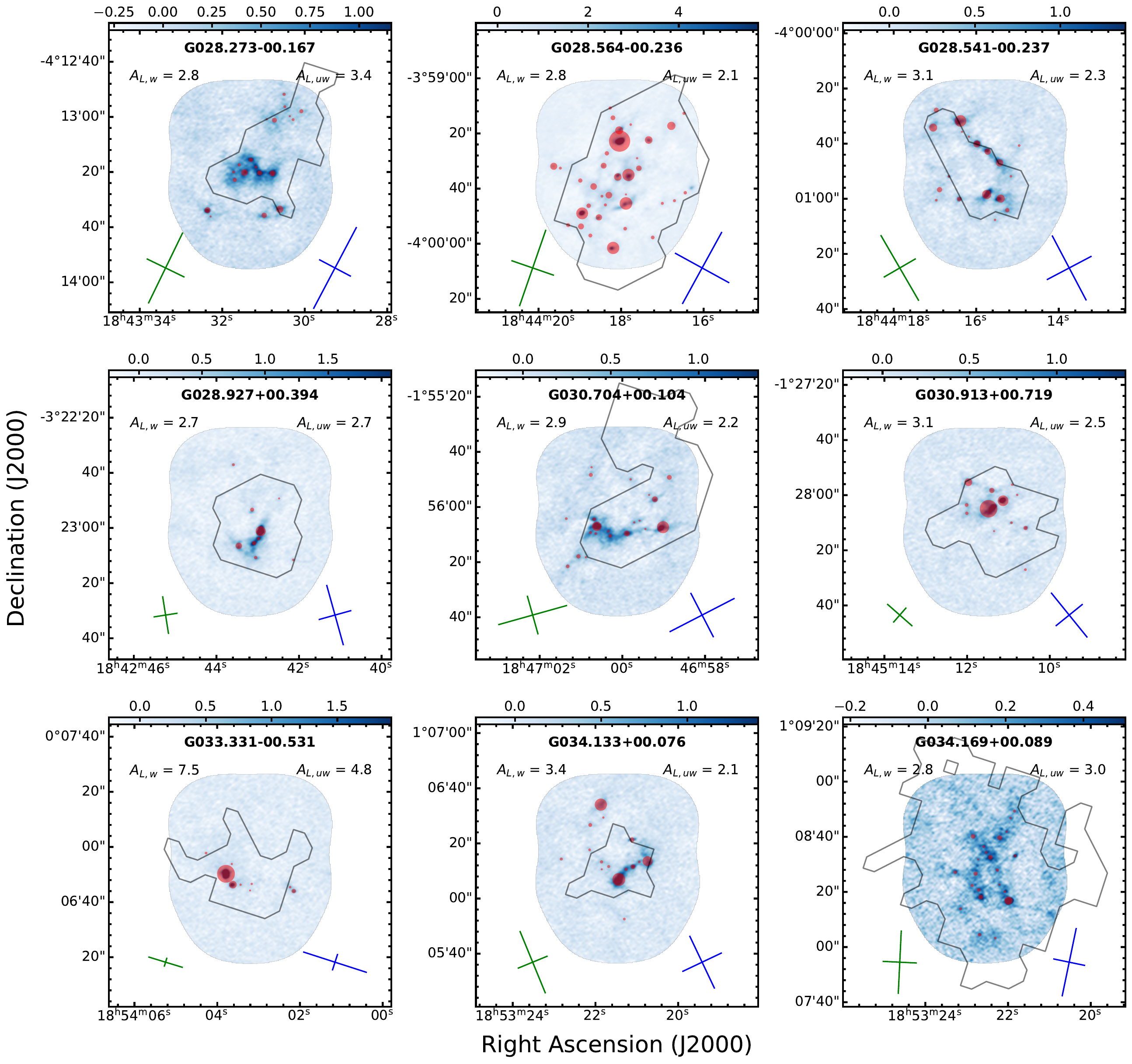}
	\caption{Continuation of Figure~\ref{fig:ASHESAL_0}. }
	\label{fig:ASHESAL_1}
\end{figure*}

\begin{figure*}[htb!]
	\epsscale{1.1}
	\plotone{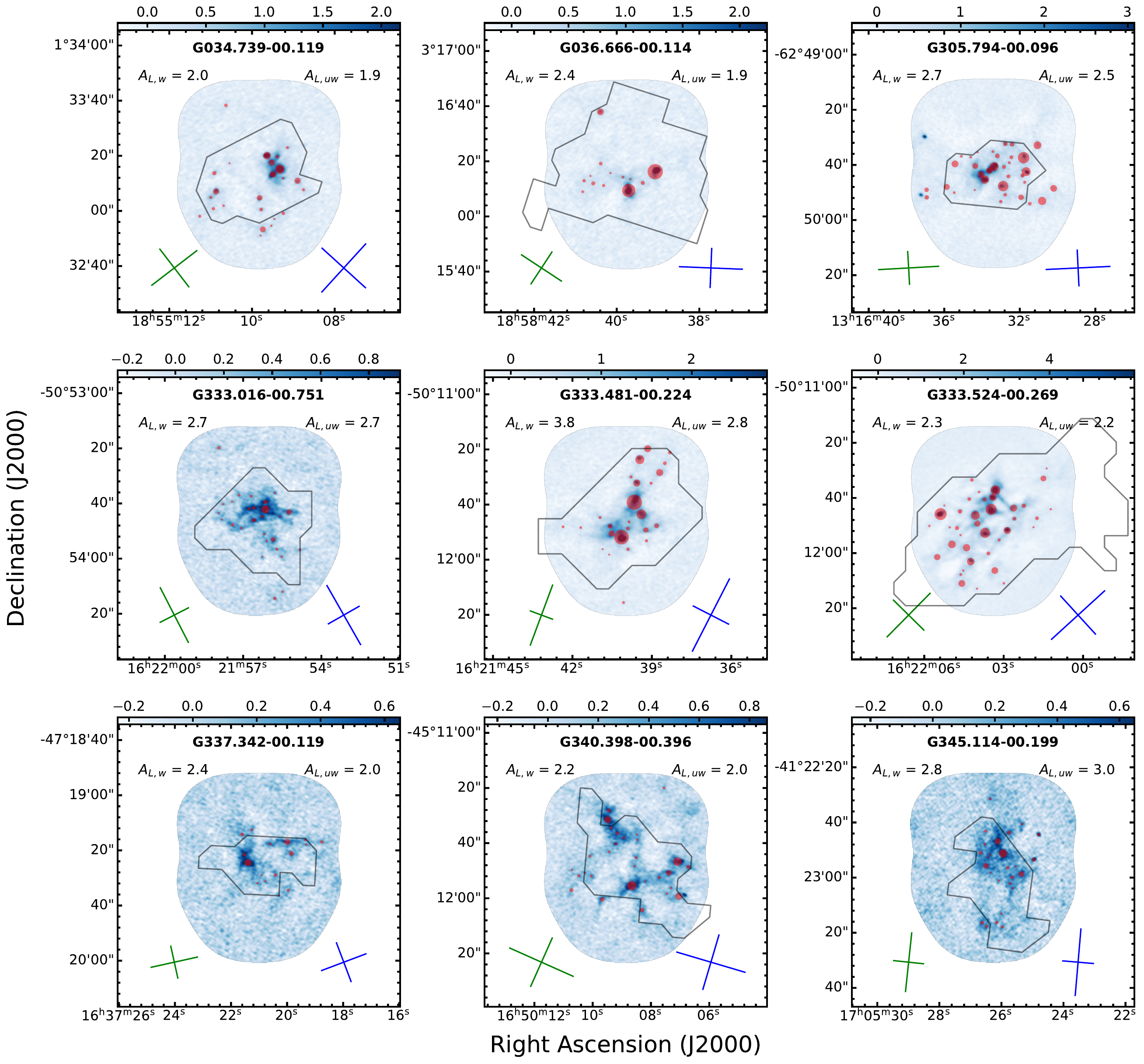}
	\caption{Continuation of Figure~\ref{fig:ASHESAL_0}. }
	\label{fig:ASHESAL_2}
\end{figure*}

\begin{figure*}[htb!]
	\epsscale{1.1}
	\plotone{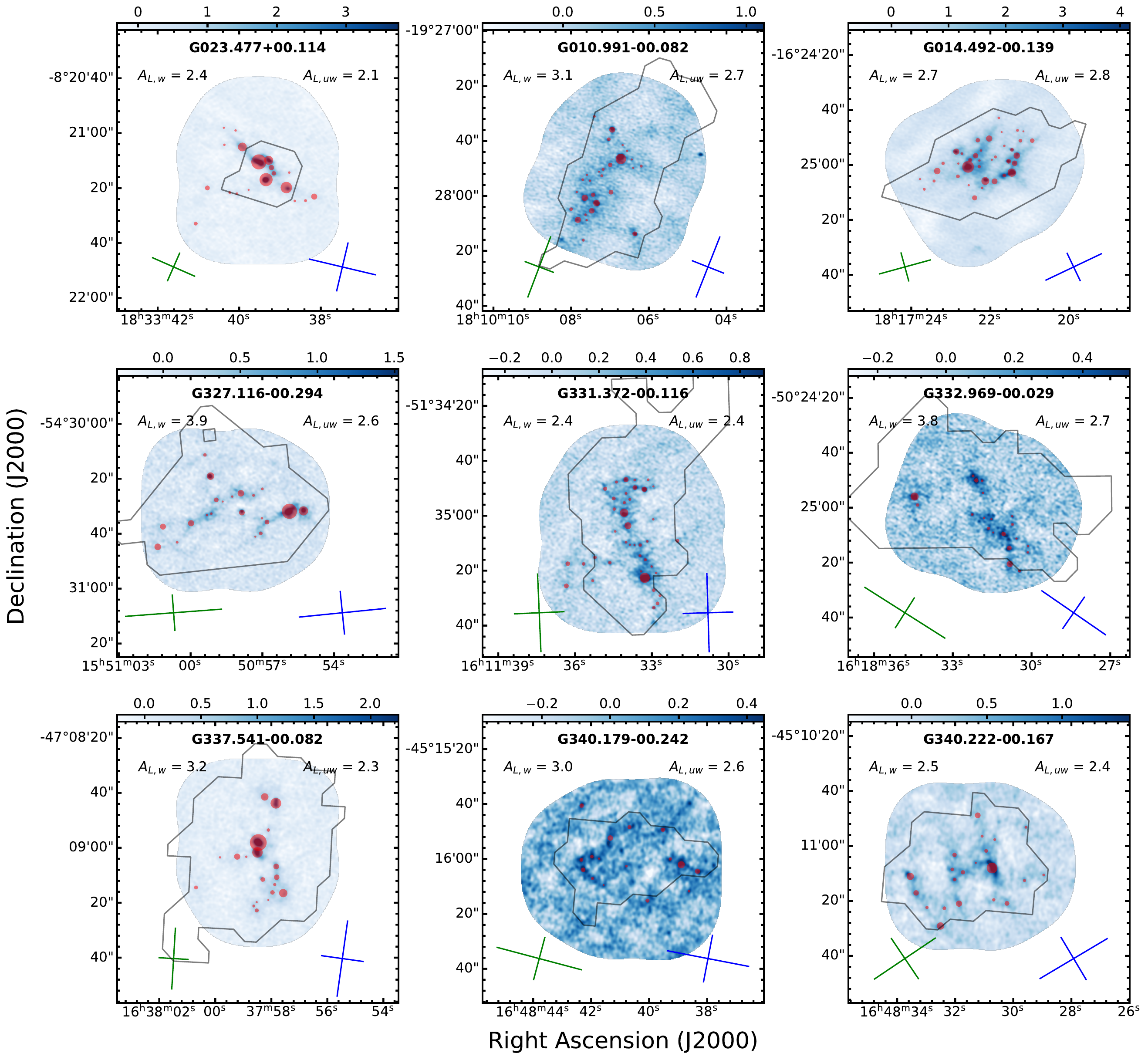}
	\caption{Continuation of Figure~\ref{fig:ASHESAL_0}. }
	\label{fig:ASHESAL_3}
\end{figure*}

\begin{figure*}[htb!]
	\epsscale{1.1}
	\plotone{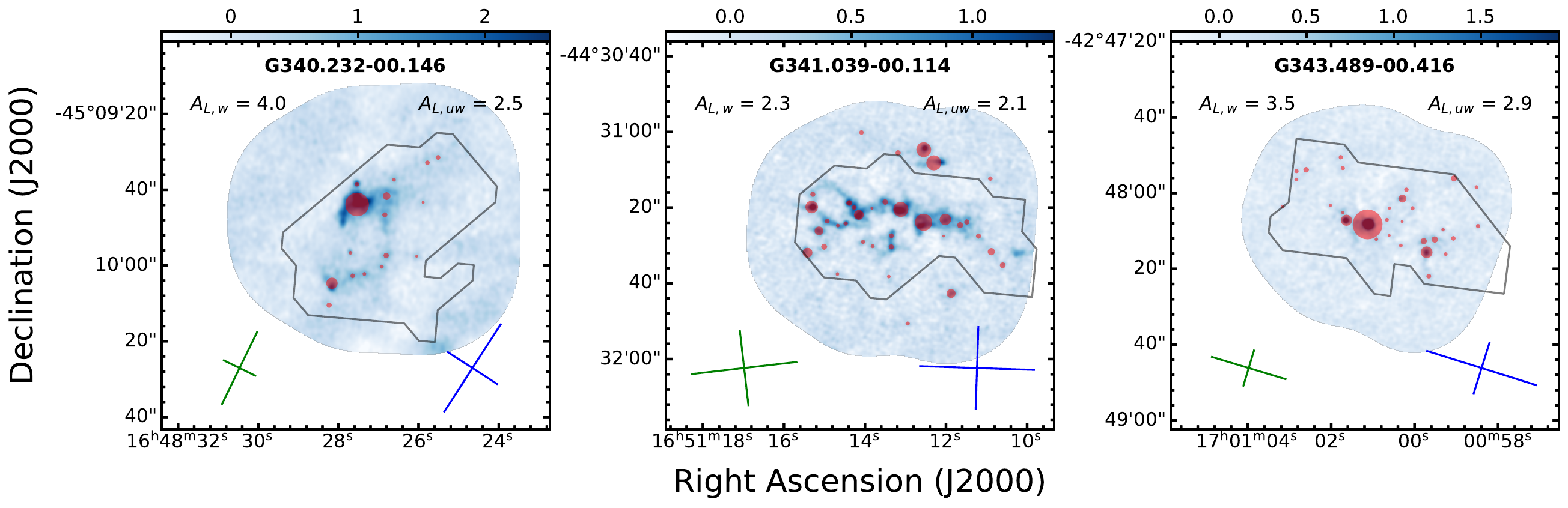}
	\caption{Continuation of Figure~\ref{fig:ASHESAL_0}. }
	\label{fig:ASHESAL_4}
\end{figure*}

\section{Influence of Uncertainty}\label{app:Uncertainty}
The uncertainties associated with each quantity analyzed in Sections~\ref{sec:CorrFrag} to~\ref{sec:Segregation} are propagated from the following estimated values: $20\%$ for temperature ($T_{cl}$), $10\%$ for distance, $50\%$ for averaged number density ($n_{cl}$), and $50\%$ for both clump mass ($M_{cl}$) and core mass. 
These uncertainties are detailed and derived in \cite{2023ApJ...950..148M} and \cite{2017ApJ...841...97S}.
For luminosity, the uncertainty originates from the combined effects of temperature, distance, and column density according to Equation 3 of \cite{2017MNRAS.466..340C}. 
A mean value of $60\%$ is then adopted (K.M., private communication).
The uncertainties for the alignment parameters are calculated from their respective errors in the mean, while the uncertainties for $\Lambda_{\text{MSR}}$ are discussed in Equation 1 of \cite{2009MNRAS.395.1449A}.
The uncertainties in $R_1$ and $R_2$ are considered negligible due to the high signal-to-noise ratio and the robust identification of structures through dendrogram at multiple noise levels.
Finally, the uncertainty in the C$^{18}$O ($J = 2-1$) velocity dispersion ($\sigma_{cl, v}$) primarily stems from the fitting of averaged line profiles using a 1D Gaussian and is estimated to be less than $10\%$.
To propagate uncertainties, Monte Carlo simulations are conducted assuming Gaussian distributions for each parameter. 
However, for parameters with significant uncertainties, such as mass and luminosity, a log-normal distribution is used to ensure positive values.

After constructing distributions for the derived parameters, new samples are generated from these distributions. 
Kendall's rank correlation tests are then performed on these samples. 
The resulting $\tau$ values and statistics are summarized in Table~\ref{tab:TauPUncertainty}.

This table presents the median and mean (with standard deviation) values for the $\tau$ value, including the one using nominal value in Table~\ref{tab:TauPNominal} for comparison. 
Based on these results, all correlations are still weak (absolute $\tau <$ 0.219) after considering parameter uncertainties. 

\begin{deluxetable*}{@{\extracolsep{4pt}}lcccccc@{}}
	\tablecaption{Statistical results of the correlation analysis using Kendall's rank correlation and considering parameter uncertainties. \label{tab:TauPUncertainty}} 
	\tablehead{
		\colhead{Parameter} & \multicolumn{3}{c}{\ALuw} & \multicolumn{3}{c}{\ALw} \\
		\cline{2-4} \cline{5-7}
		\colhead{}  & \multicolumn{2}{c}{Monte Carlo}  & \colhead{Nominal Values} & \multicolumn{2}{c}{Monte Carlo} & \colhead{Nominal Values} \\
		\cline{2-3} \cline{5-6}
		\colhead{} & \colhead{median} & \colhead{mean} & \colhead{} & \colhead{median} & \colhead{mean} & \colhead{} 
	}
	\startdata
	$\delta_{sep, avg}/\lambda_{J, cl}^{th}$ & -0.001 & -0.002 $\pm$ 0.074 & 0.012  & -0.066 & -0.067 $\pm$ 0.071 & -0.074  \\
	$\delta_{sep, avg}/\lambda_{J, cl}^{tur}$ & -0.031 & -0.031 $\pm$ 0.067 & -0.034  & -0.126 & -0.126 $\pm$ 0.061 & -0.179  \\
	Core Number & -0.097 & -0.096 $\pm$ 0.031 & -0.095 & -0.254 & -0.253 $\pm$ 0.018 & -0.254 \\
	$M_{cl}$ & -0.031 & -0.030 $\pm$ 0.055 & -0.062 & 0.082 & 0.082 $\pm$ 0.053 & 0.082 \\
	$L_{cl}$ & -0.088 & -0.089 $\pm$ 0.056 & -0.096 & -0.007 & -0.006 $\pm$ 0.053 & 0.007 \\
	$R_{cl}$ & 0.012 & 0.011 $\pm$ 0.041 & -0.020 & 0.209 & 0.210 $\pm$ 0.036 & 0.217 \\
	$T_{cl}$ & -0.080 & -0.080 $\pm$ 0.089 & -0.112 & -0.074 & -0.074 $\pm$ 0.090 & -0.087 \\
	$n_{cl}$ & -0.036 & -0.035 $\pm$ 0.065 & -0.026 & -0.158 & -0.158 $\pm$ 0.063 & -0.220 \\
	$\sigma_{cl, v}$ & -0.007 & -0.007 $\pm$ 0.050 & -0.016 & 0.061 & 0.061 $\pm$ 0.045 & 0.089 \\
	$\alpha_{vir}$ & 0.028 & 0.029 $\pm$ 0.074 & -0.005 & 0.026 & 0.026 $\pm$ 0.072 & 0.003 \\
	$R_1$ & 0.099 & 0.100 $\pm$ 0.028 & 0.107 & 0.047 & 0.047 $\pm$ 0.018 & 0.053 \\
	$R_2$ & -0.031 & -0.033 $\pm$ 0.023 & -0.023 & -0.090 & -0.091 $\pm$ 0.015 & -0.088 \\
	CFE & -0.055 & -0.054 $\pm$ 0.075 & -0.077 & -0.163 & -0.164 $\pm$ 0.074 & -0.255 \\
	$f$(proto) & -0.014 & -0.013 $\pm$ 0.031 & 0.009 & 0.028 & 0.026 $\pm$ 0.020 & 0.028 \\
	$L/M$ & -0.093 & -0.093 $\pm$ 0.070 & -0.107 & -0.080 & -0.080 $\pm$ 0.070 & -0.063  \\
	\hline \hline
	\multicolumn{1}{c}{Parameter} & \multicolumn{3}{c}{\DALAL}  \\
	\cline{2-4}
	\colhead{}  & \multicolumn{2}{c}{Monte Carlo}  & \colhead{Nominal Values} \\
	\cline{2-3}
	$\Lambda_{\text{MSR}}$ & 0.090 & 0.089 $\pm$ 0.084 & 0.136 \\
	\enddata
	\tablecomments{For a sample size of 39 data points without ties, a Kendall's $\tau$ value of 0.219 is required to achieve a $95\%$ confidence level, while $\tau$ values of 0.287 and 0.366 are needed for $99\%$ and $99.9\%$ confidence levels, respectively.}
\end{deluxetable*}

\bibliography{main}{}
\bibliographystyle{aasjournal}

\end{document}